\documentclass[
reprint,
superscriptaddress,
amsmath,amssymb,
pra,
]{revtex4-2}

\usepackage{graphicx}
\usepackage{dcolumn}
\usepackage{bm}
\usepackage{physics}
\usepackage{braket}
\usepackage{placeins}
\usepackage[colorlinks=true]{hyperref}
\usepackage{nicefrac,xfrac}
\usepackage{upgreek}

\newlength{\onecolfig}
\newlength{\twocolfig}
\newcommand{\ion}[2]{\mbox{$^{#2}$#1$^+$}}
\newcommand{\Ca}[1]{\ion{Ca}{#1}}

\newcommand{\unit}[1]{\,\mbox{#1}}

\newcommand{\Hz}{\unit{Hz}}
\newcommand{\kHz}{\unit{kHz}}
\newcommand{\MHz}{\unit{MHz}}

\newcommand{\mW}{\unit{mW}}

\newcommand{\um}{\unit{\textmu m}}

\newcommand{\K}{\unit{K}}

\newcommand{\ms}{\unit{ms}}
\newcommand{\us}{\unit{\textmu s}}

\newcommand{\mT}{\unit{mT}}

\begin{document}

\let\oldaddcontentsline\addcontentsline
\renewcommand{\addcontentsline}[3]{}

\title{Robust and fast microwave-driven quantum logic for trapped-ion qubits}

\author{M.\,A.\,Weber}
\affiliation{Clarendon Laboratory, Department of Physics, University of Oxford, Parks Road, Oxford OX1 3PU, U.K.}

\author{M.\,F.\,Gely}
\affiliation{Clarendon Laboratory, Department of Physics, University of Oxford, Parks Road, Oxford OX1 3PU, U.K.}

\author{R.\,K.\,Hanley}
\affiliation{Clarendon Laboratory, Department of Physics, University of Oxford, Parks Road, Oxford OX1 3PU, U.K.}

\author{T.\,P.\,Harty}
\affiliation{Clarendon Laboratory, Department of Physics, University of Oxford, Parks Road, Oxford OX1 3PU, U.K.}

\author{A.\,D.\,Leu}
\affiliation{Clarendon Laboratory, Department of Physics, University of Oxford, Parks Road, Oxford OX1 3PU, U.K.}

\author{C.\,M.\,L{\"o}schnauer}
\affiliation{Clarendon Laboratory, Department of Physics, University of Oxford, Parks Road, Oxford OX1 3PU, U.K.}

\author{D.\,P.\,Nadlinger} 
\affiliation{Clarendon Laboratory, Department of Physics, University of Oxford, Parks Road, Oxford OX1 3PU, U.K.}

\author{D.\,M.\,Lucas} 
\email{david.lucas@physics.ox.ac.uk}
\affiliation{Clarendon Laboratory, Department of Physics, University of Oxford, Parks Road, Oxford OX1 3PU, U.K.}

\begin{abstract}
Microwave-driven logic is a promising alternative to laser control in scaling trapped-ion based quantum processors.
We implement  M{\o}lmer-S{\o}rensen two-qubit gates on \Ca{43} hyperfine clock qubits in a cryogenic ($\approx$25\K) surface trap, driven by near-field microwaves.
We achieve gate durations of 154\us\ (with 1.0(2)\% error) and 331\us\ (0.5(1)\% error), which approaches the performance of typical laser-driven gates.
In the 331\us\ gate, we demonstrate a Walsh-modulated dynamical decoupling scheme which suppresses errors due to fluctuations in the qubit frequency as well as imperfections in the decoupling drive itself. 
\end{abstract}\date{\today}
\maketitle

\begin{figure}
	\centering
	\includegraphics[width=0.45\textwidth]{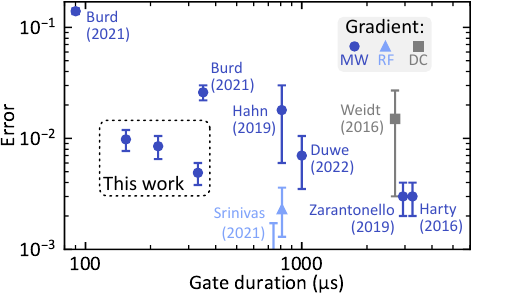}
	\caption{
		\textbf{State-of-the-art for laser-free entangling gates.}
		Bell-state preparation errors and durations for a selection of two-qubit gates demonstrated with laser-free schemes.
		Different colors correspond to different methods of engineering spatial gradients: DC magnetic field gradients or ``magic'' scheme~\cite{Weidt2016}, RF gradient~\cite{Srinivas2021}, or near-field MW gradient (as in this work)~\cite{Harty2016, Hahn2019a, Zarantonello2019, Burd2021, Duwe2022}.
	}
	\label{fig:fig1_fidelities}
\end{figure}

Quantum logic gates capable of reliably entangling qubits are a key to quantum technologies such as atomic clocks~\cite{Meyer2001}, quantum networking~\cite{Kimble2008} and quantum information processors~\cite{DiVincenzo2000}.
Trapped-ion based systems currently define the state-of-the-art in many aspects of these technologies~\cite{Brewer2019,Oelker2019,Moses2023,Nadlinger2022}.
Usually, lasers are used to drive entangling gates, with the lowest reported gate errors at the 0.1\% level, and corresponding gate durations in the 1.6\us\ to 100\us\ range~\cite{Ballance2016, Gaabler2016, Schafer2018, Clark2021}.
Alternatively, quantum gates can be driven by electronic methods, with spatial magnetic field gradients generated using direct currents (DC)~\cite{Mintert2001,Khromova2012}, radio-frequency (RF) currents~\cite{Sutherland2019,Srinivas2019} or microwave (MW) currents~\cite{Ospelkaus2008,Ospelkaus2011}.
These laser-free approaches benefit from exceptional single-qubit control and ion addressability~\cite{Harty2014,Leu2023}, the ability to embed waveguides into micro-fabricated ion traps, and the absence of photon scattering errors~\cite{Ballance2016,Moore2023}.
Compared to laser systems, MW technology also boasts lower cost, greater reliability and easier control over phase and amplitude, making MW driven logic an attractive route to scaling ion traps to larger systems.
However, MW entangling gates, at speeds similar to laser-driven gates, have not been demonstrated with errors sufficiently low for error correction~\cite{Fowler2012}.
In this Letter, we demonstrate MW driven gates with durations close to laser gates, while maintaining errors below the $\approx1\%$ error correction threshold~\cite{Fowler2012}.
As shown in Fig.~\ref{fig:fig1_fidelities}, these gate operations are an order of magnitude faster than the previous state-of-the-art for low-error gates using MW gradients~\cite{Harty2016}, five times faster than RF-gradient gates~\cite{Srinivas2021}, and have significantly lower error compared with the fastest laser-free gates~\cite{Burd2021}.
We obtain this speedup through the use of a low ion height (40\um) in a cryogenically-operated surface trap, as well as the choice of a high-field qubit ``clock'' transition in \Ca{43} at 28.8\mT.
We also propose a novel approach to dynamical decoupling, where the qubit is resonantly driven during the gate to ensure protection against qubit-frequency drifts, but the phase of the resonant drive follows a Walsh sequence to ensure no net qubit rotation results from dynamical decoupling by the end of the gate.
Such schemes have previously been considered for single-qubit gates~\cite{Ball2015}.

\begin{figure}
	\centering
	\includegraphics[width=0.45\textwidth]{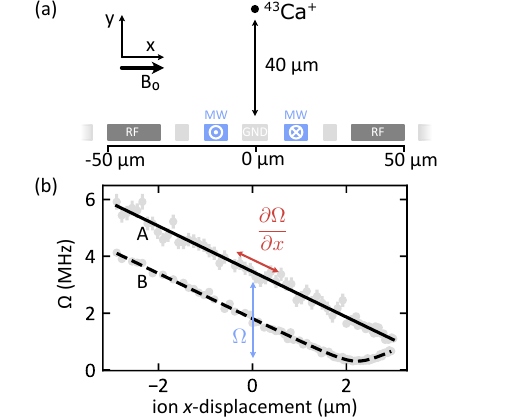}
	\caption{
		\textbf{Ion trap and field distribution.}
		(a) Cross-section of the electrode layout of the surface-electrode trap.
		A static magnetic field (labelled $B_0$) of 28.8\mT\ sets the quantization axis for the trapped $^{43}\text{Ca}^+$ ion.
		The qubit transition of the ion is driven by MW fields parallel to the quantization axis.
		This MW field component is measured and displayed in (b) as a Rabi-frequency $\Omega$ scaled to the MW power used in gate implementations.
		The Rabi-frequency is measured as a function of ion displacement in the in-plane radial direction $x$.
		The spatial gradient $\partial\Omega/\partial x$ is proportional to the gate Rabi-frequency $\Omega_g$, which sets the gate duration.
		The two MW field distributions shown, labelled A and B, correspond to measurements on two different dates, and reflect the change in trap properties attributed to structural changes in the trap upon thermal contraction when cryogenically cooling the trap to $\approx 25\K$.
	}
	\label{fig:fig2_setup}
\end{figure}

\begin{figure}
	\centering
	\includegraphics[width=0.5\textwidth]{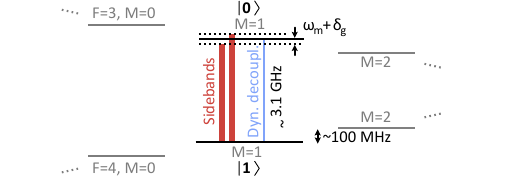}
	\caption{
		\textbf{Clock qubit transition.}
		Relevant $^{43}\text{Ca}^+$ ground level hyperfine states, and the MW dynamical decoupling and sideband tones used to drive the gate.
		Compared to previous laser-free gate demonstrations in $^{43}\text{Ca}^+$~\cite{Harty2016}, this ``$\pi$'' qubit transition benefits from a more efficient coupling to the linearly-polarized MW field, and a larger matrix element~\cite{Weber2022}, contributing, with the trap design, to the measured decrease in gate time.
	}
	\label{fig:fig3_ion}
\end{figure}

Experiments are performed in a micro-fabricated segmented-electrode surface Paul trap, operated cryogenically at $\approx25~\K$, with a ~40~\um\ ion height.
A cross-section of the electrode layout is shown in Fig.~\ref{fig:fig2_setup}(a).
The magnetic field strength generated by the counter-propagating MW currents, corresponding to the two arms of an on-chip $\uplambda/4$-resonator, is shown in Fig.~\ref{fig:fig2_setup}(b).
Upon cryogenically cycling the system, the trap has exhibited different MW field distributions, labelled A and B, which appear to be reproducible.
We suspect this is due to structural changes induced by thermal contractions, as we have previously observed similar effects due to small changes in the on-chip RF path (see~\cite{Weber2022}).
Both field distributions are featured in the different gate measurements discussed below.
A more detailed description of the trap design is provided in Ref.~\cite{Weber2022}.
A description of the control system and the MW drive chain can be found in Refs.~\cite{Leu2023,Weber2023}.
As a qubit, we use the $\ket{\text{F} =4,\text{M} =1 }$ and $\ket{\text{F} =3,\text{M}=1}$ hyperfine states within the $4\text{S}_{1/2}$ ground level manifold of $^{43}\text{Ca}^+$.
The qubit transition, as well as neighboring states, are shown in Fig.~\ref{fig:fig3_ion}.
At our operating static magnetic field of 28.8\mT, the frequency of this transition is first-order insensitive to the static magnetic field, forming a ``clock'' transition.
When compared to previous work~\cite{Harty2016}, this $\pi$-polarised qubit enables $\approx 2$ times faster two-qubit gate operations when driven by the same field.
This arises from a larger matrix element and more efficient use of the linearly-polarised MW gradients which are generated by our electrode geometry, see Ref.~\cite{Weber2022}.
The techniques used for ion state-preparation and measurement (SPAM), loading and cooling are described in Ref.~\cite{Weber2022}.
Entangling gates are mediated by a collective motional mode of the ions.
In our case we use the in-plane radial ``rocking'' (``out-of-phase'') mode of a two-ion crystal (referred to hereafter as the gate mode).
This mode was chosen for its low heating rate $\lesssim 2$~quanta/s, limiting gate errors to $\lesssim 2\times10^{-4}$.
To reduce the gate sensitivity to motional frequency fluctuations~\cite{Sutherland2022}, we use dark resonance cooling of all modes followed by Raman sideband cooling to the ground state of the gate mode, with an average number of phonons $\lesssim 0.02$.
Similarly, we ground-state cool the in-plane radial center-of-mass mode to an average number of phonons $\lesssim 0.1$, to reduce the impact of sideband dynamics driven on this spectator mode.

\begin{figure*}
	\centering
	\includegraphics[width=0.9\textwidth]{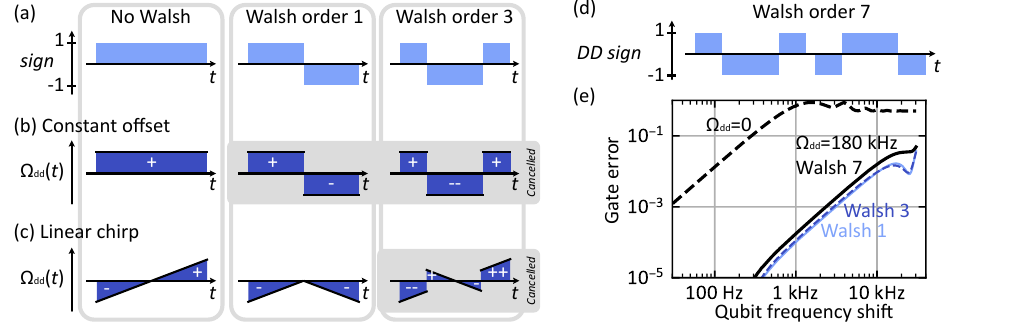}
	\caption{
		\textbf{Walsh modulated dynamical decoupling}
		(a) Sign changes for Walsh sequences of order 0 (no Walsh), 1 and 3.
		The total sequence is of the same duration as an entangling gate.
		(b) Case where the continuous dynamical decoupling Rabi-frequency $\Omega_\text{dd}$ remains constant throughout the gate.
		The qubit rotation driven by the decoupling tone (\textit{i.e.} the integral of $\Omega_\text{dd}(t)$ over time) is cancelled when Walsh 1 or 3 sequences are used.
		(c) In the case of a linear change in $\Omega_\text{dd}(t)$, a Walsh 3 sequence is required to cancel the qubit rotation (whilst still protecting against constant offsets in $\Omega_\text{dd}$).
		(d) Walsh 7 sequence used in the experiment, which additionally cancels quadratic changes in $\Omega_\text{dd}(t)$.
		(e) Numerical simulation of the gate error ($t_g = 331\us$, $N=2$) in the presence of a static qubit-frequency shift.
		Using a $\Omega_\text{dd}=180\kHz$ dynamical decoupling tone (with Walsh-7 modulation) significantly increases the gate's robustness to this error source.
	}
	\label{fig:fig4_walsh}
\end{figure*}

\begin{figure}
	\centering
	\includegraphics[width=0.45\textwidth]{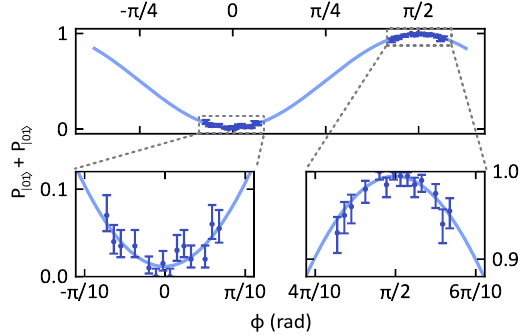}
	\caption{
		\textbf{Example of Bell-state preparation error measurement ($217\us$ gate).}
		Following standard tomography techniques~\cite{Leibfried2003}, the error in preparing a Bell-state $(\ket{00}+\ket{11})/\sqrt{2}$ is inferred from the population of its constituent states $P_{\ket{00}}$ and $P_{\ket{11}}$, measured after a gate, and a measurement of coherence, which is shown here.
		Data points show a measurement of $P_{\ket{01}}+ P_{\ket{10}}$ after subjecting the qubits to an entangling gate and a global $\pi/2$ rotation pulse with a varying phase $\phi$.
		Coherence is measured as the ability to reproducibly rotate the Bell-state into $(\ket{01}+\ket{10})/\sqrt{2}$ ($\phi=\pi/2$) or leave it unchanged ($\phi=0$) and hence corresponds to the contrast $C$ of the oscillations in the presented dataset.
		We concentrate data acquisition around $\phi=0$ and $\phi=\pi/2$ in order to maximize the information content for a given data acquisition time.
		After a maximum-likelihood-estimate fit to the data (solid line), and subtraction of the independently measured SPAM error, the Bell-state preparation error is given by $1-(C+P_{\ket{00}}$ + $P_{\ket{11}})/2=0.85(20)\%$.
	}
	\label{fig:fig5_parity_scan}
\end{figure}

The two-qubit gates in this work are of the M{\o}lmer-S{\o}rensen (MS) type~\cite{Molmer1999}.
When driving the ion crystal with a single MW tone, the Rabi Hamiltonian is modulated by the ions' motion, following
\begin{align}
\begin{split}
	\hat H =& \hbar\omega_\text{m}\hat a^\dagger\hat a+\frac{\hbar\omega_q}{2}\hat S_z\\
	&+\hbar\left(\Omega \hat S_{x,+}+\frac{1}{\sqrt{2}}\frac{\partial\Omega}{\partial x}\hat x\hat S_{x,-}\right)\cos(\omega t)\ ,
\end{split}
\label{eq:drive_Hamiltonian}
\end{align}
where $\Omega$ is the Rabi-frequency, $\omega$, $\omega_q$ and $\omega_\text{m}$ are the frequencies of the MW field, qubit transition, and motional mode, respectively.
The displacement operator for the ion motion is $\hat x=x_\text{zpf}\left(\hat a+\hat a^\dagger\right)$, with mode annihilation operator $\hat a$ and zero-point fluctuations $x_\text{zpf}=\sqrt{\hbar/2m\omega_\text{m}}$, where $m$ is the mass of a single ion.
The components of the motional mode eigenvectors appear as a factor $1/\sqrt{2}$~\cite{James1998}\footnote{We neglect in Eq.~(\ref{eq:drive_Hamiltonian}) the small $\approx 15^\circ$ angle between the field gradient and the motional mode axis.}.
The Pauli operators for the internal states of both ions (labelled 1 and 2) are combined into single expressions $\hat S_{x,\pm} = \hat\sigma_{x,1}\pm\hat\sigma_{x,2}$ and $\hat S_z = \hat\sigma_{z,1}+\hat\sigma_{z,2}$.
%
%
In the presence of two tones, with frequencies near-resonant with the red and blue motional sidebands of the qubit transition $\omega=\omega_q\pm(\omega_\text{m}+\delta_\text{g})$, and of equal amplitude (Rabi-frequency $\Omega$), the Hamiltonian approximates to
\begin{align}
	\begin{split}
\hat{H}_\text{g}&\simeq\frac{\hbar\Omega_\text{g}}{2}\hat{S}_{x,-}\left(\hat{a}e^{-i\delta_\text{g} t}+\hat{a}^{\dagger}e^{i\delta_\text{g} t}\right),\\
 \Omega_\text{g}&=x_\text{zpf}\frac{\partial\Omega}{\partial x}
\end{split}
\label{eq:MS_Hamiltonian}
\end{align}
in the interaction picture.
We will refer to $\Omega_\text{g}$ and $\delta_\text{g}$ as the gate Rabi-frequency and detuning respectively.
Under the conditions $2\Omega_\text{g}\sqrt{N}=\delta_g$ and $t_\text{g}\delta_\text{g}=2\pi N$, driving the sidebands for the gate duration $t_\text{g}$ will implement a $N\in\mathbb{N}$ loop MS gate.
The dynamics of an MS gate are disturbed by the introduction of a miscalibrated, or fluctuating, a.c.\ Zeeman shift $\Delta$ described by the Hamiltonian $\hat{H}_\text{z}=\hbar\Delta\hat S_z/2$.
Such errors can be suppressed by the introduction of a resonant MW tone~\cite{Bermudez2012,Harty2016} implementing continuous dynamical decoupling, with Hamiltonian
\begin{equation}
	\hat{H}_\text{dd}=\pm\frac{\hbar\Omega_\text{dd}}{2}\hat S_{x,+}
	\label{eq:Carrier_Hamiltonian}
\end{equation}
Since $\hat{H}_\text{dd}$ commutes with the gate Hamiltonian $\hat{H}_\text{g}$, it will not disturb the gate dynamics, but may lead to an undesired qubit rotation by the end of the gate.
To avoid this, one could use a calibrated Rabi-frequency $\Omega_\text{dd}$ such that the gate time $t_g=2M\pi/\Omega_\text{dd}$ with $M$ an integer.
However, such an approach is sensitive to Rabi-frequency fluctuations.
Harty \textit{et al.} avoid this constraint by inserting a $\pi_y$-pulse half way through the gate interaction~\cite{Harty2016}.
This method requires however that the Rabi-frequency remains constant throughout the gate.
Also, such pulses may only be inserted when the MS-interaction closes a motional phase space loop, potentially forcing the use of a slower multi-loop gate.
Here we introduce an alternative method to cancel this undesired qubit rotation, by switching the sign of the dynamical decoupling drive according to a Walsh sequence.
We note that Eq.~(\ref{eq:Carrier_Hamiltonian}) indicates a free choice of sign, corresponding to a $\pi$ phase-shift of the decoupling tone.
As both choices for $\hat{H}_\text{dd}$ commute with $\pm\hat{H}_\text{g}$, we may modulate the sign of the decoupling tone throughout the MS-gate without disturbing the two-qubit gate dynamics.
A first-order Walsh sequence corresponds to using two decoupling pulses with opposite signs.
Any rotation accumulated in the first pulse is then cancelled by the second, provided the drive strength $\Omega_\text{dd}$ is constant.
To protect against both constant and linear variations of $\Omega_\text{dd}$ throughout the gate, a third-order sequence is required, as illustrated in Figs.~\ref{fig:fig4_walsh}(a-c).
A Walsh sequence of order 7 protects against quadratic changes, order 15 addresses cubic changes, etc.
We first report on the implementation of a single-loop and two-loop MS gate, with durations $154\us$ and $217\us$ respectively, without recourse to dynamical decoupling.
The two sidebands are each driven with $\approx3.3$~W drive tones, this power being measured at the entrance to the vacuum chamber.
Based on the gate times achieved, we infer that this power generates a microwave field gradient of 89 T/m.
We achieve a gate Rabi-frequency of $\Omega_\text{g}=2\pi\times3.3\kHz$; this was accompanied by off-resonant driving of the qubit with Rabi-frequency $\Omega=2\pi\times1.81\MHz$.
At these powers, $\Omega$ is then comparable to the sideband detuning, which is close to the motional mode frequency $\omega_\text{m}=2\pi \times4.0\MHz$.
To adiabatically suppress off-resonantly driven qubit rotations, we ramp the sideband amplitude following a $\sin^2$ shape over a rise-/fall-time $2.8 \us\gg 2\pi/\sqrt{\omega_\text{m}^2+\Omega^2}$.
Thermal transients of the trap properties and MW chain were minimised by maintaining the injected MW energy per shot constant, using dummy MW pulses (even when the apparatus is idle), see Supplementary Information.
We measure the error in preparing Bell-states with these gates using standard tomography~\cite{Leibfried2003}.
After correcting for the independently measured SPAM error of $0.12(1)\%$ per ion, the single- and two-loop MS gates were measured to produce Bell states with errors of $0.98(21)\%$ and $0.85(20)\%$ respectively.
A parity measurement for the $217\us$ gate is shown in Fig.~\ref{fig:fig5_parity_scan}.
We attribute the majority of the error to drifts of the motional frequency over the course of the gate characterization.
Extrapolating from motional frequency measurements during the gate calibration measurements, we estimate that the motional mode drifted by $\sim 390 \Hz$ over $\sim 2$ minutes, giving rise to an error of $\sim0.7\%$.
The second significant error ($0.17\%$) arises from Kerr-coupling between the gate mode and the out-of-plane radial rocking mode~\cite{Nie2009}.
Indeed, due to a small projection of the out-of-plane direction on the dark-state cooling laser beam paths ($\theta=$75$^{\circ}$), the out-of-plane radial rocking mode remains thermally populated with approximately 20 phonons during the gate.
This thermal phonon distribution couples to the inter-ion spacing thereby causing uncertainty in the gate-mode frequency.
A hundredfold reduction in this error is obtained in the $331\us$ gate discussed below by reducing the axial center-of-mass mode frequency from $2.2$ to $1.1$\MHz.
However, this change brings the in-plane center-of-mass mode frequency closer to the gate mode frequency, creating an error of similar magnitude.

We secondly report on the implementation of a two-loop MS gate, with duration $331\us$, this time utilising Walsh-modulated dynamical decoupling.
Whilst qubit-frequency fluctuations are not expected to dominate the gate error, we aim to demonstrate here that the improved robustness offered by the scheme can be obtained without impacting the measured gate error.
For the resonant decoupling tone, we use a $6 \mW$ signal, producing a Rabi-frequency $\Omega_\text{dd} = 2\pi \times 180\kHz$, and which is modulated following the Walsh-7 sequence shown in Fig.~\ref{fig:fig4_walsh}(d).
Sign changes are implemented by ramping down the resonant tone and subsequently ramping it up with a modified phase over the course of 0.24\us.
From gate simulations with this pulse scheme, dynamical decoupling provides protection from a wide band of qubit-frequency shifts, as shown in Fig.~\ref{fig:fig4_walsh}(e), which may come from drifts or miscalibrations of the a.c.\ Zeeman shift generated by the sideband drive.
The a.c.\ Zeeman shift is calibrated to be $\Delta= 2\pi\times46\kHz$ in the $154\us$ and $217 \us$ gate, and $\Delta= 2\pi\times26\kHz$ in this $331 \us$ gate.
Other differences with the faster gates include the MW field distribution (A rather than B, see Fig.~\ref{fig:fig2_setup}(b)), injected power (1.8~W per sideband, corresponding to 70 T/m and a gate Rabi frequency of $\Omega_\text{g}=2\pi\times2.1\kHz$), gate mode frequency (5.6\MHz), and pulse rise/fall-time (1\us).

We measure a Bell state preparation error of $0.49(11)\%$ for this $331\us$ gate, after correcting for the SPAM error of $0.28(2)\%$ per qubit.
Again, we attribute the majority of the observed error to drifts of the gate mode frequency.
Extrapolating mode frequency measurements acquired during the gate calibration suggest a $\sim 130\Hz$ drift over the gate data acquisition, corresponding to an error of $0.2\%$.
The second significant error (0.16\%) comes from driving the sideband of a spectator motional mode, namely the in-plane radial center-of-mass mode with frequency $\omega_m+2\pi\times 120$ kHz.
This error could be mitigated by increasing the sideband tone ramp time to $\gtrsim 10\us$.
For all gates measured in this work, other known error sources were estimated to be in the low $10^{-4}$ regime or below, amounting to a negligible contribution~\cite{Weber2023}.

In conclusion, we have demonstrated microwave-driven entangling gates with gate times of 154\us, 217\us\ and 331\us\ and Bell-state preparation errors of 0.98(21)\%, 0.85(20)\% and 0.49(11)\% respectively.
This work shows that microwave-driven gates can be as fast as laser-driven gates, whilst keeping gate errors below the $\approx 1\%$ error correction threshold.
The dominant gate errors are technical in nature, and we expect that by engineering a more stable motional mode frequency, by speeding up both gate calibration and characterization~\cite{Gaebler2012,Gerster2022}, or by also using Walsh modulation in the sideband drive~\cite{Hayes2012}, the error could be lowered by an order of magnitude or more.
Additionally, we have proposed and implemented a dynamical decoupling scheme for two-qubit gate operations which is robust to qubit frequency fluctuations and to imperfections of the decoupling drive itself.
We note that this Walsh-modulated dynamical decoupling may also be used in conjunction with Walsh modulation of the sign of the sideband drive, allowing for gates which are robust against fluctuations of the microwave Rabi-, the qubit-, and the motional mode-frequency.
This scheme may also find application in laser-driven gates for suppressing errors induced by a.c.~Stark shifts~\cite{Haffner2003} and in other use cases for continuous dynamical decoupling~\cite{Smith2023}.
We note that microwave-driven gates with similar speeds, using a DC field gradient, have recently been demonstrated at the University of Siegen~\cite{nunnerich2024fast}; and RF-driven gates with similar speeds, and lower errors, have very recently been reported by Oxford Ionics Ltd~\cite{loschnauer2024}.

\vspace{5mm}
\textbf{Acknowledgements}
This work was supported by the U.S. Army Research Office (ref.\ W911NF-18-1-0340) and the U.K.\ EPSRC Quantum Computing and Simulation Hub (Ref. EP/T001062/1).
M.F.G.\ acknowledges support from the Netherlands Organization for Scientific Research (NWO) through a Rubicon Grant.
T.P.H.\ is a Director of Oxford Ionics Ltd, which partly supports A.D.L.

\appendix

\section*{Appendix: Microwave power management protocol}

\begin{figure}[h]
	\centering
	\includegraphics[width=0.45\textwidth]{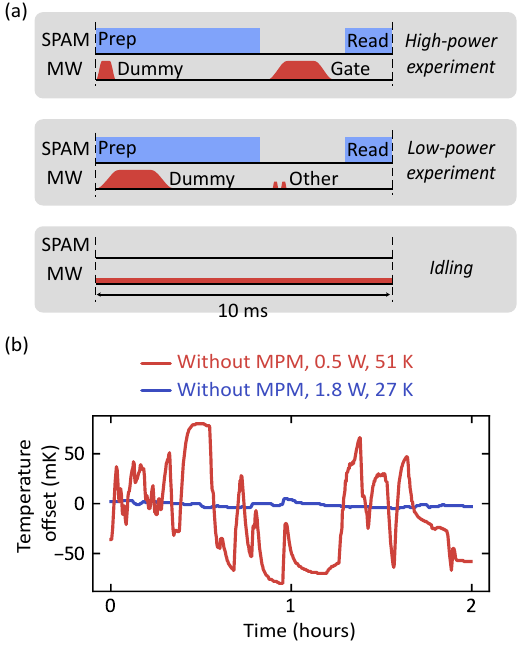}
	\caption{
		\textbf{Microwave power management (MPM).}
        (a) Schematic of the MPM protocol in two example experiments, and during idling.
        The schematic shows the state-preparation and measurement steps (blue), and microwave pulses (red) injected over the course of a shot (fixed to a 10 ms duration).
        (b) Temperature variations over time for an arbitrary sequence of high- or low-power experiments, and idling periods, with and without microwave power management.
        The success of the MPM is even more striking when considering that in the dataset with MPM approximately four times more microwave power is used ($\approx 1.8$ W per sideband rather than $\approx 0.5$ W), and the temperature is significantly lower (27 K rather than 51 K), when compared to the dataset without MPM.
	}
	\label{fig:S_duty_cycle}
\end{figure}

The high microwave power necessary to drive fast gates causes significant fluctuations in the trap temperature, a problem which is exacerbated at cryogenic temperatures.
This results in fluctuating microwave properties of the trap, which has numerous consequences for entangling gates and their characterisation.
For example, in the absence of any microwave power management, the microwave transfer pulses used for state-preparation, readout, and tomography, will fluctuate in their amplitude, introducing percent-level errors in the characterisation of the gates.
In this appendix, we summarize the microwave power management protocol used during the calibration and characterisation of entangling gates presented in this work.
Further details regarding this technique can be found in Ref.~\cite{Weber2023}.
We employ a microwave power management (MPM) technique summarized in Fig.~\ref{fig:S_duty_cycle}(a).
First, we fix the duration of each data acquisition cycle, or ``shot'', to 10 \ms.
A shot typically comprises of a state-preparation and cooling step, microwave-driven manipulations of the qubit state, and readout.
At the beginning of each shot, we use a ``dummy'' microwave pulse which ensures that the total microwave energy injected per shot is kept to a pre-defined constant value.
We apply this dummy pulse during optical state-preparation, where the internal state of the ion is pumped away from the qubit states, and thus the dummy pulses (applied with the same amplitude and frequencies as an entangling gate) have no effect on the qubits.
When the system is idling, the microwaves are kept on, again at the same frequencies as an entangling gate, however at a smaller amplitude, to maintain the average injected power constant.
As shown in Fig.~\ref{fig:S_duty_cycle}(b), this substantially reduced thermal fluctuations.
\vspace{10pt}

\FloatBarrier
\bibliography{library}

\begin{thebibliography}{45}%
\makeatletter
\providecommand \@ifxundefined [1]{%
 \@ifx{#1\undefined}
}%
\providecommand \@ifnum [1]{%
 \ifnum #1\expandafter \@firstoftwo
 \else \expandafter \@secondoftwo
 \fi
}%
\providecommand \@ifx [1]{%
 \ifx #1\expandafter \@firstoftwo
 \else \expandafter \@secondoftwo
 \fi
}%
\providecommand \natexlab [1]{#1}%
\providecommand \enquote  [1]{``#1''}%
\providecommand \bibnamefont  [1]{#1}%
\providecommand \bibfnamefont [1]{#1}%
\providecommand \citenamefont [1]{#1}%
\providecommand \href@noop [0]{\@secondoftwo}%
\providecommand \href [0]{\begingroup \@sanitize@url \@href}%
\providecommand \@href[1]{\@@startlink{#1}\@@href}%
\providecommand \@@href[1]{\endgroup#1\@@endlink}%
\providecommand \@sanitize@url [0]{\catcode `\\12\catcode `\$12\catcode
  `\&12\catcode `\#12\catcode `\^12\catcode `\_12\catcode `\%12\relax}%
\providecommand \@@startlink[1]{}%
\providecommand \@@endlink[0]{}%
\providecommand \url  [0]{\begingroup\@sanitize@url \@url }%
\providecommand \@url [1]{\endgroup\@href {#1}{\urlprefix }}%
\providecommand \urlprefix  [0]{URL }%
\providecommand \Eprint [0]{\href }%
\providecommand \doibase [0]{https://doi.org/}%
\providecommand \selectlanguage [0]{\@gobble}%
\providecommand \bibinfo  [0]{\@secondoftwo}%
\providecommand \bibfield  [0]{\@secondoftwo}%
\providecommand \translation [1]{[#1]}%
\providecommand \BibitemOpen [0]{}%
\providecommand \bibitemStop [0]{}%
\providecommand \bibitemNoStop [0]{.\EOS\space}%
\providecommand \EOS [0]{\spacefactor3000\relax}%
\providecommand \BibitemShut  [1]{\csname bibitem#1\endcsname}%
\let\auto@bib@innerbib\@empty
\bibitem [{\citenamefont {Meyer}\ \emph {et~al.}(2001)\citenamefont {Meyer},
  \citenamefont {Rowe}, \citenamefont {Kielpinski}, \citenamefont {Sackett},
  \citenamefont {Itano}, \citenamefont {Monroe},\ and\ \citenamefont
  {Wineland}}]{Meyer2001}%
  \BibitemOpen
  \bibfield  {author} {\bibinfo {author} {\bibfnamefont {V.}~\bibnamefont
  {Meyer}}, \bibinfo {author} {\bibfnamefont {M.~A.}\ \bibnamefont {Rowe}},
  \bibinfo {author} {\bibfnamefont {D.}~\bibnamefont {Kielpinski}}, \bibinfo
  {author} {\bibfnamefont {C.~A.}\ \bibnamefont {Sackett}}, \bibinfo {author}
  {\bibfnamefont {W.~M.}\ \bibnamefont {Itano}}, \bibinfo {author}
  {\bibfnamefont {C.}~\bibnamefont {Monroe}},\ and\ \bibinfo {author}
  {\bibfnamefont {D.~J.}\ \bibnamefont {Wineland}},\ }\bibfield  {title}
  {\bibinfo {title} {{Experimental demonstration of entanglement-enhanced
  rotation angle estimation using trapped ions}},\ }\href
  {https://doi.org/10.1103/PhysRevLett.86.5870} {\bibfield  {journal} {\bibinfo
   {journal} {Phys. Rev. Lett.}\ }\textbf {\bibinfo {volume} {86}},\ \bibinfo
  {pages} {5870} (\bibinfo {year} {2001})}\BibitemShut {NoStop}%
\bibitem [{\citenamefont {Kimble}(2008)}]{Kimble2008}%
  \BibitemOpen
  \bibfield  {author} {\bibinfo {author} {\bibfnamefont {H.~J.}\ \bibnamefont
  {Kimble}},\ }\bibfield  {title} {\bibinfo {title} {The quantum internet},\
  }\href {https://doi.org/10.1038/nature07127} {\bibfield  {journal} {\bibinfo
  {journal} {Nature}\ }\textbf {\bibinfo {volume} {453}},\ \bibinfo {pages}
  {1023} (\bibinfo {year} {2008})}\BibitemShut {NoStop}%
\bibitem [{\citenamefont {DiVincenzo}(2000)}]{DiVincenzo2000}%
  \BibitemOpen
  \bibfield  {author} {\bibinfo {author} {\bibfnamefont {D.~P.}\ \bibnamefont
  {DiVincenzo}},\ }\bibfield  {title} {\bibinfo {title} {{The physical
  implementation of quantum computation}},\ }\href
  {https://doi.org/10.1002/1521-3978(200009)48:9/11<771::AID-PROP771>3.0.CO;2-E}
  {\bibfield  {journal} {\bibinfo  {journal} {Fortschritte der Physik}\
  }\textbf {\bibinfo {volume} {48}},\ \bibinfo {pages} {771} (\bibinfo {year}
  {2000})},\ \Eprint {https://arxiv.org/abs/0002077} {0002077} \BibitemShut
  {NoStop}%
\bibitem [{\citenamefont {Brewer}\ \emph {et~al.}(2019)\citenamefont {Brewer},
  \citenamefont {Chen}, \citenamefont {Hankin}, \citenamefont {Clements},
  \citenamefont {Chou}, \citenamefont {Wineland}, \citenamefont {Hume},\ and\
  \citenamefont {Leibrandt}}]{Brewer2019}%
  \BibitemOpen
  \bibfield  {author} {\bibinfo {author} {\bibfnamefont {S.~M.}\ \bibnamefont
  {Brewer}}, \bibinfo {author} {\bibfnamefont {J.-S.}\ \bibnamefont {Chen}},
  \bibinfo {author} {\bibfnamefont {A.~M.}\ \bibnamefont {Hankin}}, \bibinfo
  {author} {\bibfnamefont {E.~R.}\ \bibnamefont {Clements}}, \bibinfo {author}
  {\bibfnamefont {C.~W.}\ \bibnamefont {Chou}}, \bibinfo {author}
  {\bibfnamefont {D.~J.}\ \bibnamefont {Wineland}}, \bibinfo {author}
  {\bibfnamefont {D.~B.}\ \bibnamefont {Hume}},\ and\ \bibinfo {author}
  {\bibfnamefont {D.~R.}\ \bibnamefont {Leibrandt}},\ }\bibfield  {title}
  {\bibinfo {title} {$^{27}al^{+}$ quantum-logic clock with a systematic
  uncertainty below ${10}^{-18}$},\ }\href
  {https://doi.org/10.1103/PhysRevLett.123.033201} {\bibfield  {journal}
  {\bibinfo  {journal} {Phys. Rev. Lett.}\ }\textbf {\bibinfo {volume} {123}},\
  \bibinfo {pages} {033201} (\bibinfo {year} {2019})}\BibitemShut {NoStop}%
\bibitem [{\citenamefont {Oelker}\ \emph {et~al.}(2019)\citenamefont {Oelker},
  \citenamefont {Hutson}, \citenamefont {Kennedy}, \citenamefont {Sonderhouse},
  \citenamefont {Bothwell}, \citenamefont {Goban}, \citenamefont {Kedar},
  \citenamefont {Sanner}, \citenamefont {Robinson}, \citenamefont {Marti},
  \citenamefont {Matei}, \citenamefont {Legero}, \citenamefont {Giunta},
  \citenamefont {Holzwarth}, \citenamefont {Riehle}, \citenamefont {Sterr},\
  and\ \citenamefont {Ye}}]{Oelker2019}%
  \BibitemOpen
  \bibfield  {author} {\bibinfo {author} {\bibfnamefont {E.}~\bibnamefont
  {Oelker}}, \bibinfo {author} {\bibfnamefont {R.~B.}\ \bibnamefont {Hutson}},
  \bibinfo {author} {\bibfnamefont {C.~J.}\ \bibnamefont {Kennedy}}, \bibinfo
  {author} {\bibfnamefont {L.}~\bibnamefont {Sonderhouse}}, \bibinfo {author}
  {\bibfnamefont {T.}~\bibnamefont {Bothwell}}, \bibinfo {author}
  {\bibfnamefont {A.}~\bibnamefont {Goban}}, \bibinfo {author} {\bibfnamefont
  {D.}~\bibnamefont {Kedar}}, \bibinfo {author} {\bibfnamefont
  {C.}~\bibnamefont {Sanner}}, \bibinfo {author} {\bibfnamefont {J.~M.}\
  \bibnamefont {Robinson}}, \bibinfo {author} {\bibfnamefont {G.~E.}\
  \bibnamefont {Marti}}, \bibinfo {author} {\bibfnamefont {D.~G.}\ \bibnamefont
  {Matei}}, \bibinfo {author} {\bibfnamefont {T.}~\bibnamefont {Legero}},
  \bibinfo {author} {\bibfnamefont {M.}~\bibnamefont {Giunta}}, \bibinfo
  {author} {\bibfnamefont {R.}~\bibnamefont {Holzwarth}}, \bibinfo {author}
  {\bibfnamefont {F.}~\bibnamefont {Riehle}}, \bibinfo {author} {\bibfnamefont
  {U.}~\bibnamefont {Sterr}},\ and\ \bibinfo {author} {\bibfnamefont
  {J.}~\bibnamefont {Ye}},\ }\bibfield  {title} {\bibinfo {title}
  {Demonstration of 4.8$\times 10^{-17}$ stability at 1s for two independent
  optical clocks},\ }\href {https://doi.org/10.1038/s41566-019-0493-4}
  {\bibfield  {journal} {\bibinfo  {journal} {Nat. Photonics}\ }\textbf
  {\bibinfo {volume} {13}},\ \bibinfo {pages} {714} (\bibinfo {year}
  {2019})}\BibitemShut {NoStop}%
\bibitem [{\citenamefont {Moses}\ \emph {et~al.}(2023)\citenamefont {Moses},
  \citenamefont {Baldwin}, \citenamefont {Allman}, \citenamefont {Ancona},
  \citenamefont {Ascarrunz}, \citenamefont {Barnes}, \citenamefont
  {Bartolotta}, \citenamefont {Bjork}, \citenamefont {Blanchard}, \citenamefont
  {Bohn}, \citenamefont {Bohnet}, \citenamefont {Brown}, \citenamefont
  {Burdick}, \citenamefont {Burton}, \citenamefont {Campbell}, \citenamefont
  {Campora}, \citenamefont {Carron}, \citenamefont {Chambers}, \citenamefont
  {Chan}, \citenamefont {Chen}, \citenamefont {Chernoguzov}, \citenamefont
  {Chertkov}, \citenamefont {Colina}, \citenamefont {Curtis}, \citenamefont
  {Daniel}, \citenamefont {DeCross}, \citenamefont {Deen}, \citenamefont
  {Delaney}, \citenamefont {Dreiling}, \citenamefont {Ertsgaard}, \citenamefont
  {Esposito}, \citenamefont {Estey}, \citenamefont {Fabrikant}, \citenamefont
  {Figgatt}, \citenamefont {Foltz}, \citenamefont {Foss-Feig}, \citenamefont
  {Francois}, \citenamefont {Gaebler}, \citenamefont {Gatterman}, \citenamefont
  {Gilbreth}, \citenamefont {Giles}, \citenamefont {Glynn}, \citenamefont
  {Hall}, \citenamefont {Hankin}, \citenamefont {Hansen}, \citenamefont
  {Hayes}, \citenamefont {Higashi}, \citenamefont {Hoffman}, \citenamefont
  {Horning}, \citenamefont {Hout}, \citenamefont {Jacobs}, \citenamefont
  {Johansen}, \citenamefont {Jones}, \citenamefont {Karcz}, \citenamefont
  {Klein}, \citenamefont {Lauria}, \citenamefont {Lee}, \citenamefont {Liefer},
  \citenamefont {Lu}, \citenamefont {Lucchetti}, \citenamefont {Lytle},
  \citenamefont {Malm}, \citenamefont {Matheny}, \citenamefont {Mathewson},
  \citenamefont {Mayer}, \citenamefont {Miller}, \citenamefont {Mills},
  \citenamefont {Neyenhuis}, \citenamefont {Nugent}, \citenamefont {Olson},
  \citenamefont {Parks}, \citenamefont {Price}, \citenamefont {Price},
  \citenamefont {Pugh}, \citenamefont {Ransford}, \citenamefont {Reed},
  \citenamefont {Roman}, \citenamefont {Rowe}, \citenamefont {Ryan-Anderson},
  \citenamefont {Sanders}, \citenamefont {Sedlacek}, \citenamefont {Shevchuk},
  \citenamefont {Siegfried}, \citenamefont {Skripka}, \citenamefont {Spaun},
  \citenamefont {Sprenkle}, \citenamefont {Stutz}, \citenamefont {Swallows},
  \citenamefont {Tobey}, \citenamefont {Tran}, \citenamefont {Tran},
  \citenamefont {Vogt}, \citenamefont {Volin}, \citenamefont {Walker},
  \citenamefont {Zolot},\ and\ \citenamefont {Pino}}]{Moses2023}%
  \BibitemOpen
  \bibfield  {author} {\bibinfo {author} {\bibfnamefont {S.~A.}\ \bibnamefont
  {Moses}}, \bibinfo {author} {\bibfnamefont {C.~H.}\ \bibnamefont {Baldwin}},
  \bibinfo {author} {\bibfnamefont {M.~S.}\ \bibnamefont {Allman}}, \bibinfo
  {author} {\bibfnamefont {R.}~\bibnamefont {Ancona}}, \bibinfo {author}
  {\bibfnamefont {L.}~\bibnamefont {Ascarrunz}}, \bibinfo {author}
  {\bibfnamefont {C.}~\bibnamefont {Barnes}}, \bibinfo {author} {\bibfnamefont
  {J.}~\bibnamefont {Bartolotta}}, \bibinfo {author} {\bibfnamefont
  {B.}~\bibnamefont {Bjork}}, \bibinfo {author} {\bibfnamefont
  {P.}~\bibnamefont {Blanchard}}, \bibinfo {author} {\bibfnamefont
  {M.}~\bibnamefont {Bohn}}, \bibinfo {author} {\bibfnamefont {J.~G.}\
  \bibnamefont {Bohnet}}, \bibinfo {author} {\bibfnamefont {N.~C.}\
  \bibnamefont {Brown}}, \bibinfo {author} {\bibfnamefont {N.~Q.}\ \bibnamefont
  {Burdick}}, \bibinfo {author} {\bibfnamefont {W.~C.}\ \bibnamefont {Burton}},
  \bibinfo {author} {\bibfnamefont {S.~L.}\ \bibnamefont {Campbell}}, \bibinfo
  {author} {\bibfnamefont {J.~P.}\ \bibnamefont {Campora}}, \bibinfo {author}
  {\bibfnamefont {C.}~\bibnamefont {Carron}}, \bibinfo {author} {\bibfnamefont
  {J.}~\bibnamefont {Chambers}}, \bibinfo {author} {\bibfnamefont {J.~W.}\
  \bibnamefont {Chan}}, \bibinfo {author} {\bibfnamefont {Y.~H.}\ \bibnamefont
  {Chen}}, \bibinfo {author} {\bibfnamefont {A.}~\bibnamefont {Chernoguzov}},
  \bibinfo {author} {\bibfnamefont {E.}~\bibnamefont {Chertkov}}, \bibinfo
  {author} {\bibfnamefont {J.}~\bibnamefont {Colina}}, \bibinfo {author}
  {\bibfnamefont {J.~P.}\ \bibnamefont {Curtis}}, \bibinfo {author}
  {\bibfnamefont {R.}~\bibnamefont {Daniel}}, \bibinfo {author} {\bibfnamefont
  {M.}~\bibnamefont {DeCross}}, \bibinfo {author} {\bibfnamefont
  {D.}~\bibnamefont {Deen}}, \bibinfo {author} {\bibfnamefont {C.}~\bibnamefont
  {Delaney}}, \bibinfo {author} {\bibfnamefont {J.~M.}\ \bibnamefont
  {Dreiling}}, \bibinfo {author} {\bibfnamefont {C.~T.}\ \bibnamefont
  {Ertsgaard}}, \bibinfo {author} {\bibfnamefont {J.}~\bibnamefont {Esposito}},
  \bibinfo {author} {\bibfnamefont {B.}~\bibnamefont {Estey}}, \bibinfo
  {author} {\bibfnamefont {M.}~\bibnamefont {Fabrikant}}, \bibinfo {author}
  {\bibfnamefont {C.}~\bibnamefont {Figgatt}}, \bibinfo {author} {\bibfnamefont
  {C.}~\bibnamefont {Foltz}}, \bibinfo {author} {\bibfnamefont
  {M.}~\bibnamefont {Foss-Feig}}, \bibinfo {author} {\bibfnamefont
  {D.}~\bibnamefont {Francois}}, \bibinfo {author} {\bibfnamefont {J.~P.}\
  \bibnamefont {Gaebler}}, \bibinfo {author} {\bibfnamefont {T.~M.}\
  \bibnamefont {Gatterman}}, \bibinfo {author} {\bibfnamefont {C.~N.}\
  \bibnamefont {Gilbreth}}, \bibinfo {author} {\bibfnamefont {J.}~\bibnamefont
  {Giles}}, \bibinfo {author} {\bibfnamefont {E.}~\bibnamefont {Glynn}},
  \bibinfo {author} {\bibfnamefont {A.}~\bibnamefont {Hall}}, \bibinfo {author}
  {\bibfnamefont {A.~M.}\ \bibnamefont {Hankin}}, \bibinfo {author}
  {\bibfnamefont {A.}~\bibnamefont {Hansen}}, \bibinfo {author} {\bibfnamefont
  {D.}~\bibnamefont {Hayes}}, \bibinfo {author} {\bibfnamefont
  {B.}~\bibnamefont {Higashi}}, \bibinfo {author} {\bibfnamefont {I.~M.}\
  \bibnamefont {Hoffman}}, \bibinfo {author} {\bibfnamefont {B.}~\bibnamefont
  {Horning}}, \bibinfo {author} {\bibfnamefont {J.~J.}\ \bibnamefont {Hout}},
  \bibinfo {author} {\bibfnamefont {R.}~\bibnamefont {Jacobs}}, \bibinfo
  {author} {\bibfnamefont {J.}~\bibnamefont {Johansen}}, \bibinfo {author}
  {\bibfnamefont {L.}~\bibnamefont {Jones}}, \bibinfo {author} {\bibfnamefont
  {J.}~\bibnamefont {Karcz}}, \bibinfo {author} {\bibfnamefont
  {T.}~\bibnamefont {Klein}}, \bibinfo {author} {\bibfnamefont
  {P.}~\bibnamefont {Lauria}}, \bibinfo {author} {\bibfnamefont
  {P.}~\bibnamefont {Lee}}, \bibinfo {author} {\bibfnamefont {D.}~\bibnamefont
  {Liefer}}, \bibinfo {author} {\bibfnamefont {S.~T.}\ \bibnamefont {Lu}},
  \bibinfo {author} {\bibfnamefont {D.}~\bibnamefont {Lucchetti}}, \bibinfo
  {author} {\bibfnamefont {C.}~\bibnamefont {Lytle}}, \bibinfo {author}
  {\bibfnamefont {A.}~\bibnamefont {Malm}}, \bibinfo {author} {\bibfnamefont
  {M.}~\bibnamefont {Matheny}}, \bibinfo {author} {\bibfnamefont
  {B.}~\bibnamefont {Mathewson}}, \bibinfo {author} {\bibfnamefont
  {K.}~\bibnamefont {Mayer}}, \bibinfo {author} {\bibfnamefont {D.~B.}\
  \bibnamefont {Miller}}, \bibinfo {author} {\bibfnamefont {M.}~\bibnamefont
  {Mills}}, \bibinfo {author} {\bibfnamefont {B.}~\bibnamefont {Neyenhuis}},
  \bibinfo {author} {\bibfnamefont {L.}~\bibnamefont {Nugent}}, \bibinfo
  {author} {\bibfnamefont {S.}~\bibnamefont {Olson}}, \bibinfo {author}
  {\bibfnamefont {J.}~\bibnamefont {Parks}}, \bibinfo {author} {\bibfnamefont
  {G.~N.}\ \bibnamefont {Price}}, \bibinfo {author} {\bibfnamefont
  {Z.}~\bibnamefont {Price}}, \bibinfo {author} {\bibfnamefont
  {M.}~\bibnamefont {Pugh}}, \bibinfo {author} {\bibfnamefont {A.}~\bibnamefont
  {Ransford}}, \bibinfo {author} {\bibfnamefont {A.~P.}\ \bibnamefont {Reed}},
  \bibinfo {author} {\bibfnamefont {C.}~\bibnamefont {Roman}}, \bibinfo
  {author} {\bibfnamefont {M.}~\bibnamefont {Rowe}}, \bibinfo {author}
  {\bibfnamefont {C.}~\bibnamefont {Ryan-Anderson}}, \bibinfo {author}
  {\bibfnamefont {S.}~\bibnamefont {Sanders}}, \bibinfo {author} {\bibfnamefont
  {J.}~\bibnamefont {Sedlacek}}, \bibinfo {author} {\bibfnamefont
  {P.}~\bibnamefont {Shevchuk}}, \bibinfo {author} {\bibfnamefont
  {P.}~\bibnamefont {Siegfried}}, \bibinfo {author} {\bibfnamefont
  {T.}~\bibnamefont {Skripka}}, \bibinfo {author} {\bibfnamefont
  {B.}~\bibnamefont {Spaun}}, \bibinfo {author} {\bibfnamefont {R.~T.}\
  \bibnamefont {Sprenkle}}, \bibinfo {author} {\bibfnamefont {R.~P.}\
  \bibnamefont {Stutz}}, \bibinfo {author} {\bibfnamefont {M.}~\bibnamefont
  {Swallows}}, \bibinfo {author} {\bibfnamefont {R.~I.}\ \bibnamefont {Tobey}},
  \bibinfo {author} {\bibfnamefont {A.}~\bibnamefont {Tran}}, \bibinfo {author}
  {\bibfnamefont {T.}~\bibnamefont {Tran}}, \bibinfo {author} {\bibfnamefont
  {E.}~\bibnamefont {Vogt}}, \bibinfo {author} {\bibfnamefont {C.}~\bibnamefont
  {Volin}}, \bibinfo {author} {\bibfnamefont {J.}~\bibnamefont {Walker}},
  \bibinfo {author} {\bibfnamefont {A.~M.}\ \bibnamefont {Zolot}},\ and\
  \bibinfo {author} {\bibfnamefont {J.~M.}\ \bibnamefont {Pino}},\ }\bibfield
  {title} {\bibinfo {title} {A race-track trapped-ion quantum processor},\
  }\href {https://doi.org/10.1103/PhysRevX.13.041052} {\bibfield  {journal}
  {\bibinfo  {journal} {Phys. Rev. X}\ }\textbf {\bibinfo {volume} {13}},\
  \bibinfo {pages} {041052} (\bibinfo {year} {2023})}\BibitemShut {NoStop}%
\bibitem [{\citenamefont {Nadlinger}\ \emph {et~al.}(2022)\citenamefont
  {Nadlinger}, \citenamefont {Drmota}, \citenamefont {Nichol}, \citenamefont
  {Araneda}, \citenamefont {Main}, \citenamefont {Srinivas}, \citenamefont
  {Lucas}, \citenamefont {Ballance}, \citenamefont {Ivanov}, \citenamefont
  {Tan}, \citenamefont {Sekatski}, \citenamefont {Urbanke}, \citenamefont
  {Renner}, \citenamefont {Sangouard},\ and\ \citenamefont
  {Bancal}}]{Nadlinger2022}%
  \BibitemOpen
  \bibfield  {author} {\bibinfo {author} {\bibfnamefont {D.~P.}\ \bibnamefont
  {Nadlinger}}, \bibinfo {author} {\bibfnamefont {P.}~\bibnamefont {Drmota}},
  \bibinfo {author} {\bibfnamefont {B.~C.}\ \bibnamefont {Nichol}}, \bibinfo
  {author} {\bibfnamefont {G.}~\bibnamefont {Araneda}}, \bibinfo {author}
  {\bibfnamefont {D.}~\bibnamefont {Main}}, \bibinfo {author} {\bibfnamefont
  {R.}~\bibnamefont {Srinivas}}, \bibinfo {author} {\bibfnamefont {D.~M.}\
  \bibnamefont {Lucas}}, \bibinfo {author} {\bibfnamefont {C.~J.}\ \bibnamefont
  {Ballance}}, \bibinfo {author} {\bibfnamefont {K.}~\bibnamefont {Ivanov}},
  \bibinfo {author} {\bibfnamefont {E.~Y.-Z.}\ \bibnamefont {Tan}}, \bibinfo
  {author} {\bibfnamefont {P.}~\bibnamefont {Sekatski}}, \bibinfo {author}
  {\bibfnamefont {R.~L.}\ \bibnamefont {Urbanke}}, \bibinfo {author}
  {\bibfnamefont {R.}~\bibnamefont {Renner}}, \bibinfo {author} {\bibfnamefont
  {N.}~\bibnamefont {Sangouard}},\ and\ \bibinfo {author} {\bibfnamefont
  {J.-D.}\ \bibnamefont {Bancal}},\ }\bibfield  {title} {\bibinfo {title}
  {Experimental quantum key distribution certified by bell's theorem},\ }\href
  {https://doi.org/10.1038/s41586-022-04941-5} {\bibfield  {journal} {\bibinfo
  {journal} {Nature}\ }\textbf {\bibinfo {volume} {607}},\ \bibinfo {pages}
  {682} (\bibinfo {year} {2022})}\BibitemShut {NoStop}%
\bibitem [{\citenamefont {Ballance}\ \emph {et~al.}(2016)\citenamefont
  {Ballance}, \citenamefont {Harty}, \citenamefont {Linke}, \citenamefont
  {Sepiol},\ and\ \citenamefont {Lucas}}]{Ballance2016}%
  \BibitemOpen
  \bibfield  {author} {\bibinfo {author} {\bibfnamefont {C.~J.}\ \bibnamefont
  {Ballance}}, \bibinfo {author} {\bibfnamefont {T.~P.}\ \bibnamefont {Harty}},
  \bibinfo {author} {\bibfnamefont {N.~M.}\ \bibnamefont {Linke}}, \bibinfo
  {author} {\bibfnamefont {M.~A.}\ \bibnamefont {Sepiol}},\ and\ \bibinfo
  {author} {\bibfnamefont {D.~M.}\ \bibnamefont {Lucas}},\ }\bibfield  {title}
  {\bibinfo {title} {High-fidelity quantum logic gates using trapped-ion
  hyperfine qubits},\ }\href {https://doi.org/10.1103/PhysRevLett.117.060504}
  {\bibfield  {journal} {\bibinfo  {journal} {Phys. Rev. Lett.}\ }\textbf
  {\bibinfo {volume} {117}},\ \bibinfo {pages} {060504} (\bibinfo {year}
  {2016})}\BibitemShut {NoStop}%
\bibitem [{\citenamefont {Gaebler}\ \emph {et~al.}(2016)\citenamefont
  {Gaebler}, \citenamefont {Tan}, \citenamefont {Lin}, \citenamefont {Wan},
  \citenamefont {Bowler}, \citenamefont {Keith}, \citenamefont {Glancy},
  \citenamefont {Coakley}, \citenamefont {Knill}, \citenamefont {Leibfried},\
  and\ \citenamefont {Wineland}}]{Gaabler2016}%
  \BibitemOpen
  \bibfield  {author} {\bibinfo {author} {\bibfnamefont {J.~P.}\ \bibnamefont
  {Gaebler}}, \bibinfo {author} {\bibfnamefont {T.~R.}\ \bibnamefont {Tan}},
  \bibinfo {author} {\bibfnamefont {Y.}~\bibnamefont {Lin}}, \bibinfo {author}
  {\bibfnamefont {Y.}~\bibnamefont {Wan}}, \bibinfo {author} {\bibfnamefont
  {R.}~\bibnamefont {Bowler}}, \bibinfo {author} {\bibfnamefont {A.~C.}\
  \bibnamefont {Keith}}, \bibinfo {author} {\bibfnamefont {S.}~\bibnamefont
  {Glancy}}, \bibinfo {author} {\bibfnamefont {K.}~\bibnamefont {Coakley}},
  \bibinfo {author} {\bibfnamefont {E.}~\bibnamefont {Knill}}, \bibinfo
  {author} {\bibfnamefont {D.}~\bibnamefont {Leibfried}},\ and\ \bibinfo
  {author} {\bibfnamefont {D.~J.}\ \bibnamefont {Wineland}},\ }\bibfield
  {title} {\bibinfo {title} {High-fidelity universal gate set for
  ${^{9}\mathrm{Be}}^{+}$ ion qubits},\ }\href
  {https://doi.org/10.1103/PhysRevLett.117.060505} {\bibfield  {journal}
  {\bibinfo  {journal} {Phys. Rev. Lett.}\ }\textbf {\bibinfo {volume} {117}},\
  \bibinfo {pages} {060505} (\bibinfo {year} {2016})}\BibitemShut {NoStop}%
\bibitem [{\citenamefont {Sch{\"{a}}fer}\ \emph {et~al.}(2018)\citenamefont
  {Sch{\"{a}}fer}, \citenamefont {Ballance}, \citenamefont {Thirumalai},
  \citenamefont {Stephenson}, \citenamefont {Ballance}, \citenamefont
  {Steane},\ and\ \citenamefont {Lucas}}]{Schafer2018}%
  \BibitemOpen
  \bibfield  {author} {\bibinfo {author} {\bibfnamefont {V.~M.}\ \bibnamefont
  {Sch{\"{a}}fer}}, \bibinfo {author} {\bibfnamefont {C.~J.}\ \bibnamefont
  {Ballance}}, \bibinfo {author} {\bibfnamefont {K.}~\bibnamefont
  {Thirumalai}}, \bibinfo {author} {\bibfnamefont {L.~J.}\ \bibnamefont
  {Stephenson}}, \bibinfo {author} {\bibfnamefont {T.~G.}\ \bibnamefont
  {Ballance}}, \bibinfo {author} {\bibfnamefont {A.~M.}\ \bibnamefont
  {Steane}},\ and\ \bibinfo {author} {\bibfnamefont {D.~M.}\ \bibnamefont
  {Lucas}},\ }\bibfield  {title} {\bibinfo {title} {{Fast quantum logic gates
  with trapped-ion qubits}},\ }\href {https://doi.org/10.1038/nature25737}
  {\bibfield  {journal} {\bibinfo  {journal} {Nature}\ }\textbf {\bibinfo
  {volume} {555}},\ \bibinfo {pages} {75} (\bibinfo {year} {2018})}\BibitemShut
  {NoStop}%
\bibitem [{\citenamefont {Clark}\ \emph {et~al.}(2021)\citenamefont {Clark},
  \citenamefont {Tinkey}, \citenamefont {Sawyer}, \citenamefont {Meier},
  \citenamefont {Burkhardt}, \citenamefont {Seck}, \citenamefont {Shappert},
  \citenamefont {Guise}, \citenamefont {Volin}, \citenamefont {Fallek},
  \citenamefont {Hayden}, \citenamefont {Rellergert},\ and\ \citenamefont
  {Brown}}]{Clark2021}%
  \BibitemOpen
  \bibfield  {author} {\bibinfo {author} {\bibfnamefont {C.~R.}\ \bibnamefont
  {Clark}}, \bibinfo {author} {\bibfnamefont {H.~N.}\ \bibnamefont {Tinkey}},
  \bibinfo {author} {\bibfnamefont {B.~C.}\ \bibnamefont {Sawyer}}, \bibinfo
  {author} {\bibfnamefont {A.~M.}\ \bibnamefont {Meier}}, \bibinfo {author}
  {\bibfnamefont {K.~A.}\ \bibnamefont {Burkhardt}}, \bibinfo {author}
  {\bibfnamefont {C.~M.}\ \bibnamefont {Seck}}, \bibinfo {author}
  {\bibfnamefont {C.~M.}\ \bibnamefont {Shappert}}, \bibinfo {author}
  {\bibfnamefont {N.~D.}\ \bibnamefont {Guise}}, \bibinfo {author}
  {\bibfnamefont {C.~E.}\ \bibnamefont {Volin}}, \bibinfo {author}
  {\bibfnamefont {S.~D.}\ \bibnamefont {Fallek}}, \bibinfo {author}
  {\bibfnamefont {H.~T.}\ \bibnamefont {Hayden}}, \bibinfo {author}
  {\bibfnamefont {W.~G.}\ \bibnamefont {Rellergert}},\ and\ \bibinfo {author}
  {\bibfnamefont {K.~R.}\ \bibnamefont {Brown}},\ }\bibfield  {title} {\bibinfo
  {title} {High-fidelity bell-state preparation with $^{40}{\mathrm{ca}}^{+}$
  optical qubits},\ }\href {https://doi.org/10.1103/PhysRevLett.127.130505}
  {\bibfield  {journal} {\bibinfo  {journal} {Phys. Rev. Lett.}\ }\textbf
  {\bibinfo {volume} {127}},\ \bibinfo {pages} {130505} (\bibinfo {year}
  {2021})}\BibitemShut {NoStop}%
\bibitem [{\citenamefont {Mintert}\ and\ \citenamefont
  {Wunderlich}(2001)}]{Mintert2001}%
  \BibitemOpen
  \bibfield  {author} {\bibinfo {author} {\bibfnamefont {F.}~\bibnamefont
  {Mintert}}\ and\ \bibinfo {author} {\bibfnamefont {C.}~\bibnamefont
  {Wunderlich}},\ }\bibfield  {title} {\bibinfo {title} {{Ion-trap quantum
  logic using long-wavelength radiation}},\ }\href
  {https://doi.org/10.1103/PhysRevLett.87.257904} {\bibfield  {journal}
  {\bibinfo  {journal} {Phys. Rev. Lett.}\ }\textbf {\bibinfo {volume} {87}},\
  \bibinfo {pages} {257904} (\bibinfo {year} {2001})}\BibitemShut {NoStop}%
\bibitem [{\citenamefont {Khromova}\ \emph {et~al.}(2012)\citenamefont
  {Khromova}, \citenamefont {Piltz}, \citenamefont {Scharfenberger},
  \citenamefont {Gloger}, \citenamefont {Johanning}, \citenamefont {Var\'on},\
  and\ \citenamefont {Wunderlich}}]{Khromova2012}%
  \BibitemOpen
  \bibfield  {author} {\bibinfo {author} {\bibfnamefont {A.}~\bibnamefont
  {Khromova}}, \bibinfo {author} {\bibfnamefont {C.}~\bibnamefont {Piltz}},
  \bibinfo {author} {\bibfnamefont {B.}~\bibnamefont {Scharfenberger}},
  \bibinfo {author} {\bibfnamefont {T.~F.}\ \bibnamefont {Gloger}}, \bibinfo
  {author} {\bibfnamefont {M.}~\bibnamefont {Johanning}}, \bibinfo {author}
  {\bibfnamefont {A.~F.}\ \bibnamefont {Var\'on}},\ and\ \bibinfo {author}
  {\bibfnamefont {C.}~\bibnamefont {Wunderlich}},\ }\bibfield  {title}
  {\bibinfo {title} {Designer spin pseudomolecule implemented with trapped ions
  in a magnetic gradient},\ }\href
  {https://doi.org/10.1103/PhysRevLett.108.220502} {\bibfield  {journal}
  {\bibinfo  {journal} {Phys. Rev. Lett.}\ }\textbf {\bibinfo {volume} {108}},\
  \bibinfo {pages} {220502} (\bibinfo {year} {2012})}\BibitemShut {NoStop}%
\bibitem [{\citenamefont {Sutherland}\ \emph {et~al.}(2019)\citenamefont
  {Sutherland}, \citenamefont {Srinivas}, \citenamefont {Burd}, \citenamefont
  {Leibfried}, \citenamefont {Wilson}, \citenamefont {Wineland}, \citenamefont
  {Allcock}, \citenamefont {Slichter},\ and\ \citenamefont
  {Libby}}]{Sutherland2019}%
  \BibitemOpen
  \bibfield  {author} {\bibinfo {author} {\bibfnamefont {R.~T.}\ \bibnamefont
  {Sutherland}}, \bibinfo {author} {\bibfnamefont {R.}~\bibnamefont
  {Srinivas}}, \bibinfo {author} {\bibfnamefont {S.~C.}\ \bibnamefont {Burd}},
  \bibinfo {author} {\bibfnamefont {D.}~\bibnamefont {Leibfried}}, \bibinfo
  {author} {\bibfnamefont {A.~C.}\ \bibnamefont {Wilson}}, \bibinfo {author}
  {\bibfnamefont {D.~J.}\ \bibnamefont {Wineland}}, \bibinfo {author}
  {\bibfnamefont {D.~T.~C.}\ \bibnamefont {Allcock}}, \bibinfo {author}
  {\bibfnamefont {D.~H.}\ \bibnamefont {Slichter}},\ and\ \bibinfo {author}
  {\bibfnamefont {S.~B.}\ \bibnamefont {Libby}},\ }\bibfield  {title} {\bibinfo
  {title} {Versatile laser-free trapped-ion entangling gates},\ }\href
  {https://doi.org/10.1088/1367-2630/ab0be5} {\bibfield  {journal} {\bibinfo
  {journal} {New J. Phys.}\ }\textbf {\bibinfo {volume} {21}},\ \bibinfo
  {pages} {033033} (\bibinfo {year} {2019})}\BibitemShut {NoStop}%
\bibitem [{\citenamefont {Srinivas}\ \emph {et~al.}(2019)\citenamefont
  {Srinivas}, \citenamefont {Burd}, \citenamefont {Sutherland}, \citenamefont
  {Wilson}, \citenamefont {Wineland}, \citenamefont {Leibfried}, \citenamefont
  {Allcock},\ and\ \citenamefont {Slichter}}]{Srinivas2019}%
  \BibitemOpen
  \bibfield  {author} {\bibinfo {author} {\bibfnamefont {R.}~\bibnamefont
  {Srinivas}}, \bibinfo {author} {\bibfnamefont {S.~C.}\ \bibnamefont {Burd}},
  \bibinfo {author} {\bibfnamefont {R.~T.}\ \bibnamefont {Sutherland}},
  \bibinfo {author} {\bibfnamefont {A.~C.}\ \bibnamefont {Wilson}}, \bibinfo
  {author} {\bibfnamefont {D.~J.}\ \bibnamefont {Wineland}}, \bibinfo {author}
  {\bibfnamefont {D.}~\bibnamefont {Leibfried}}, \bibinfo {author}
  {\bibfnamefont {D.~T.~C.}\ \bibnamefont {Allcock}},\ and\ \bibinfo {author}
  {\bibfnamefont {D.~H.}\ \bibnamefont {Slichter}},\ }\bibfield  {title}
  {\bibinfo {title} {Trapped-ion spin-motion coupling with microwaves and a
  near-motional oscillating magnetic field gradient},\ }\href
  {https://doi.org/10.1103/PhysRevLett.122.163201} {\bibfield  {journal}
  {\bibinfo  {journal} {Phys. Rev. Lett.}\ }\textbf {\bibinfo {volume} {122}},\
  \bibinfo {pages} {163201} (\bibinfo {year} {2019})}\BibitemShut {NoStop}%
\bibitem [{\citenamefont {Ospelkaus}\ \emph {et~al.}(2008)\citenamefont
  {Ospelkaus}, \citenamefont {Langer}, \citenamefont {Amini}, \citenamefont
  {Brown}, \citenamefont {Leibfried},\ and\ \citenamefont
  {Wineland}}]{Ospelkaus2008}%
  \BibitemOpen
  \bibfield  {author} {\bibinfo {author} {\bibfnamefont {C.}~\bibnamefont
  {Ospelkaus}}, \bibinfo {author} {\bibfnamefont {C.~E.}\ \bibnamefont
  {Langer}}, \bibinfo {author} {\bibfnamefont {J.~M.}\ \bibnamefont {Amini}},
  \bibinfo {author} {\bibfnamefont {K.~R.}\ \bibnamefont {Brown}}, \bibinfo
  {author} {\bibfnamefont {D.}~\bibnamefont {Leibfried}},\ and\ \bibinfo
  {author} {\bibfnamefont {D.~J.}\ \bibnamefont {Wineland}},\ }\bibfield
  {title} {\bibinfo {title} {Trapped-ion quantum logic gates based on
  oscillating magnetic fields},\ }\href
  {https://doi.org/10.1103/PhysRevLett.101.090502} {\bibfield  {journal}
  {\bibinfo  {journal} {Phys. Rev. Lett.}\ }\textbf {\bibinfo {volume} {101}},\
  \bibinfo {pages} {090502} (\bibinfo {year} {2008})}\BibitemShut {NoStop}%
\bibitem [{\citenamefont {Ospelkaus}\ \emph {et~al.}(2011)\citenamefont
  {Ospelkaus}, \citenamefont {Warring}, \citenamefont {Colombe}, \citenamefont
  {Brown}, \citenamefont {Amini}, \citenamefont {Leibfried},\ and\
  \citenamefont {Wineland}}]{Ospelkaus2011}%
  \BibitemOpen
  \bibfield  {author} {\bibinfo {author} {\bibfnamefont {C.}~\bibnamefont
  {Ospelkaus}}, \bibinfo {author} {\bibfnamefont {U.}~\bibnamefont {Warring}},
  \bibinfo {author} {\bibfnamefont {Y.}~\bibnamefont {Colombe}}, \bibinfo
  {author} {\bibfnamefont {K.~R.}\ \bibnamefont {Brown}}, \bibinfo {author}
  {\bibfnamefont {J.~M.}\ \bibnamefont {Amini}}, \bibinfo {author}
  {\bibfnamefont {D.}~\bibnamefont {Leibfried}},\ and\ \bibinfo {author}
  {\bibfnamefont {D.~J.}\ \bibnamefont {Wineland}},\ }\bibfield  {title}
  {\bibinfo {title} {{Microwave quantum logic gates for trapped ions}},\ }\href
  {https://doi.org/10.1038/nature10290} {\bibfield  {journal} {\bibinfo
  {journal} {Nature}\ }\textbf {\bibinfo {volume} {476}},\ \bibinfo {pages}
  {181} (\bibinfo {year} {2011})}\BibitemShut {NoStop}%
\bibitem [{\citenamefont {Harty}\ \emph {et~al.}(2014)\citenamefont {Harty},
  \citenamefont {Allcock}, \citenamefont {Ballance}, \citenamefont {Guidoni},
  \citenamefont {Janacek}, \citenamefont {Linke}, \citenamefont {Stacey},\ and\
  \citenamefont {Lucas}}]{Harty2014}%
  \BibitemOpen
  \bibfield  {author} {\bibinfo {author} {\bibfnamefont {T.~P.}\ \bibnamefont
  {Harty}}, \bibinfo {author} {\bibfnamefont {D.~T.~C.}\ \bibnamefont
  {Allcock}}, \bibinfo {author} {\bibfnamefont {C.~J.}\ \bibnamefont
  {Ballance}}, \bibinfo {author} {\bibfnamefont {L.}~\bibnamefont {Guidoni}},
  \bibinfo {author} {\bibfnamefont {H.~A.}\ \bibnamefont {Janacek}}, \bibinfo
  {author} {\bibfnamefont {N.~M.}\ \bibnamefont {Linke}}, \bibinfo {author}
  {\bibfnamefont {D.~N.}\ \bibnamefont {Stacey}},\ and\ \bibinfo {author}
  {\bibfnamefont {D.~M.}\ \bibnamefont {Lucas}},\ }\bibfield  {title} {\bibinfo
  {title} {High-fidelity preparation, gates, memory, and readout of a
  trapped-ion quantum bit},\ }\href
  {https://doi.org/10.1103/PhysRevLett.113.220501} {\bibfield  {journal}
  {\bibinfo  {journal} {Phys. Rev. Lett.}\ }\textbf {\bibinfo {volume} {113}},\
  \bibinfo {pages} {220501} (\bibinfo {year} {2014})}\BibitemShut {NoStop}%
\bibitem [{\citenamefont {Leu}\ \emph {et~al.}(2023)\citenamefont {Leu},
  \citenamefont {Gely}, \citenamefont {Weber}, \citenamefont {Smith},
  \citenamefont {Nadlinger},\ and\ \citenamefont {Lucas}}]{Leu2023}%
  \BibitemOpen
  \bibfield  {author} {\bibinfo {author} {\bibfnamefont {A.~D.}\ \bibnamefont
  {Leu}}, \bibinfo {author} {\bibfnamefont {M.~F.}\ \bibnamefont {Gely}},
  \bibinfo {author} {\bibfnamefont {M.~A.}\ \bibnamefont {Weber}}, \bibinfo
  {author} {\bibfnamefont {M.~C.}\ \bibnamefont {Smith}}, \bibinfo {author}
  {\bibfnamefont {D.~P.}\ \bibnamefont {Nadlinger}},\ and\ \bibinfo {author}
  {\bibfnamefont {D.~M.}\ \bibnamefont {Lucas}},\ }\bibfield  {title} {\bibinfo
  {title} {Fast, high-fidelity addressed single-qubit gates using efficient
  composite pulse sequences},\ }\href
  {https://doi.org/10.1103/PhysRevLett.131.120601} {\bibfield  {journal}
  {\bibinfo  {journal} {Phys. Rev. Lett.}\ }\textbf {\bibinfo {volume} {131}},\
  \bibinfo {pages} {120601} (\bibinfo {year} {2023})}\BibitemShut {NoStop}%
\bibitem [{\citenamefont {Moore}\ \emph {et~al.}(2023)\citenamefont {Moore},
  \citenamefont {Campbell}, \citenamefont {Hudson}, \citenamefont
  {Boguslawski}, \citenamefont {Wineland},\ and\ \citenamefont
  {Allcock}}]{Moore2023}%
  \BibitemOpen
  \bibfield  {author} {\bibinfo {author} {\bibfnamefont {I.~D.}\ \bibnamefont
  {Moore}}, \bibinfo {author} {\bibfnamefont {W.~C.}\ \bibnamefont {Campbell}},
  \bibinfo {author} {\bibfnamefont {E.~R.}\ \bibnamefont {Hudson}}, \bibinfo
  {author} {\bibfnamefont {M.~J.}\ \bibnamefont {Boguslawski}}, \bibinfo
  {author} {\bibfnamefont {D.~J.}\ \bibnamefont {Wineland}},\ and\ \bibinfo
  {author} {\bibfnamefont {D.~T.~C.}\ \bibnamefont {Allcock}},\ }\bibfield
  {title} {\bibinfo {title} {Photon scattering errors during stimulated raman
  transitions in trapped-ion qubits},\ }\href
  {https://doi.org/10.1103/PhysRevA.107.032413} {\bibfield  {journal} {\bibinfo
   {journal} {Phys. Rev. A}\ }\textbf {\bibinfo {volume} {107}},\ \bibinfo
  {pages} {032413} (\bibinfo {year} {2023})}\BibitemShut {NoStop}%
\bibitem [{\citenamefont {Fowler}\ \emph {et~al.}(2012)\citenamefont {Fowler},
  \citenamefont {Mariantoni}, \citenamefont {Martinis},\ and\ \citenamefont
  {Cleland}}]{Fowler2012}%
  \BibitemOpen
  \bibfield  {author} {\bibinfo {author} {\bibfnamefont {A.~G.}\ \bibnamefont
  {Fowler}}, \bibinfo {author} {\bibfnamefont {M.}~\bibnamefont {Mariantoni}},
  \bibinfo {author} {\bibfnamefont {J.~M.}\ \bibnamefont {Martinis}},\ and\
  \bibinfo {author} {\bibfnamefont {A.~N.}\ \bibnamefont {Cleland}},\
  }\bibfield  {title} {\bibinfo {title} {Surface codes: Towards practical
  large-scale quantum computation},\ }\href
  {https://doi.org/10.1103/PhysRevA.86.032324} {\bibfield  {journal} {\bibinfo
  {journal} {Phys. Rev. A}\ }\textbf {\bibinfo {volume} {86}},\ \bibinfo
  {pages} {032324} (\bibinfo {year} {2012})}\BibitemShut {NoStop}%
\bibitem [{\citenamefont {Harty}\ \emph {et~al.}(2016)\citenamefont {Harty},
  \citenamefont {Sepiol}, \citenamefont {Allcock}, \citenamefont {Ballance},
  \citenamefont {Tarlton},\ and\ \citenamefont {Lucas}}]{Harty2016}%
  \BibitemOpen
  \bibfield  {author} {\bibinfo {author} {\bibfnamefont {T.~P.}\ \bibnamefont
  {Harty}}, \bibinfo {author} {\bibfnamefont {M.~A.}\ \bibnamefont {Sepiol}},
  \bibinfo {author} {\bibfnamefont {D.~T.~C.}\ \bibnamefont {Allcock}},
  \bibinfo {author} {\bibfnamefont {C.~J.}\ \bibnamefont {Ballance}}, \bibinfo
  {author} {\bibfnamefont {J.~E.}\ \bibnamefont {Tarlton}},\ and\ \bibinfo
  {author} {\bibfnamefont {D.~M.}\ \bibnamefont {Lucas}},\ }\bibfield  {title}
  {\bibinfo {title} {High-fidelity trapped-ion quantum logic using near-field
  microwaves},\ }\href {https://doi.org/10.1103/PhysRevLett.117.140501}
  {\bibfield  {journal} {\bibinfo  {journal} {Phys. Rev. Lett.}\ }\textbf
  {\bibinfo {volume} {117}},\ \bibinfo {pages} {140501} (\bibinfo {year}
  {2016})}\BibitemShut {NoStop}%
\bibitem [{\citenamefont {Srinivas}\ \emph {et~al.}(2021)\citenamefont
  {Srinivas}, \citenamefont {Burd}, \citenamefont {Knaack}, \citenamefont
  {Sutherland}, \citenamefont {Kwiatkowski}, \citenamefont {Glancy},
  \citenamefont {Knill}, \citenamefont {Wineland}, \citenamefont {Leibfried},
  \citenamefont {Wilson}, \citenamefont {Allcock},\ and\ \citenamefont
  {Slichter}}]{Srinivas2021}%
  \BibitemOpen
  \bibfield  {author} {\bibinfo {author} {\bibfnamefont {R.}~\bibnamefont
  {Srinivas}}, \bibinfo {author} {\bibfnamefont {S.~C.}\ \bibnamefont {Burd}},
  \bibinfo {author} {\bibfnamefont {H.~M.}\ \bibnamefont {Knaack}}, \bibinfo
  {author} {\bibfnamefont {R.~T.}\ \bibnamefont {Sutherland}}, \bibinfo
  {author} {\bibfnamefont {A.}~\bibnamefont {Kwiatkowski}}, \bibinfo {author}
  {\bibfnamefont {S.}~\bibnamefont {Glancy}}, \bibinfo {author} {\bibfnamefont
  {E.}~\bibnamefont {Knill}}, \bibinfo {author} {\bibfnamefont {D.~J.}\
  \bibnamefont {Wineland}}, \bibinfo {author} {\bibfnamefont {D.}~\bibnamefont
  {Leibfried}}, \bibinfo {author} {\bibfnamefont {A.~C.}\ \bibnamefont
  {Wilson}}, \bibinfo {author} {\bibfnamefont {D.~T.~C.}\ \bibnamefont
  {Allcock}},\ and\ \bibinfo {author} {\bibfnamefont {D.~H.}\ \bibnamefont
  {Slichter}},\ }\bibfield  {title} {\bibinfo {title} {High-fidelity laser-free
  universal control of trapped ion qubits},\ }\href
  {https://doi.org/10.1038/s41586-021-03809-4} {\bibfield  {journal} {\bibinfo
  {journal} {Nature}\ }\textbf {\bibinfo {volume} {597}},\ \bibinfo {pages}
  {209} (\bibinfo {year} {2021})}\BibitemShut {NoStop}%
\bibitem [{\citenamefont {Burd}\ \emph {et~al.}(2021)\citenamefont {Burd},
  \citenamefont {Srinivas}, \citenamefont {Knaack}, \citenamefont {Ge},
  \citenamefont {Wilson}, \citenamefont {Wineland}, \citenamefont {Leibfried},
  \citenamefont {Bollinger}, \citenamefont {Allcock},\ and\ \citenamefont
  {Slichter}}]{Burd2021}%
  \BibitemOpen
  \bibfield  {author} {\bibinfo {author} {\bibfnamefont {S.~C.}\ \bibnamefont
  {Burd}}, \bibinfo {author} {\bibfnamefont {R.}~\bibnamefont {Srinivas}},
  \bibinfo {author} {\bibfnamefont {H.~M.}\ \bibnamefont {Knaack}}, \bibinfo
  {author} {\bibfnamefont {W.}~\bibnamefont {Ge}}, \bibinfo {author}
  {\bibfnamefont {A.~C.}\ \bibnamefont {Wilson}}, \bibinfo {author}
  {\bibfnamefont {D.~J.}\ \bibnamefont {Wineland}}, \bibinfo {author}
  {\bibfnamefont {D.}~\bibnamefont {Leibfried}}, \bibinfo {author}
  {\bibfnamefont {J.~J.}\ \bibnamefont {Bollinger}}, \bibinfo {author}
  {\bibfnamefont {D.~T.}\ \bibnamefont {Allcock}},\ and\ \bibinfo {author}
  {\bibfnamefont {D.~H.}\ \bibnamefont {Slichter}},\ }\bibfield  {title}
  {\bibinfo {title} {{Quantum amplification of boson-mediated interactions}},\
  }\href {https://doi.org/10.1038/s41567-021-01237-9} {\bibfield  {journal}
  {\bibinfo  {journal} {Nat. Phys}\ }\textbf {\bibinfo {volume} {17}},\
  \bibinfo {pages} {898} (\bibinfo {year} {2021})}\BibitemShut {NoStop}%
\bibitem [{\citenamefont {Ball}\ and\ \citenamefont
  {Biercuk}(2015)}]{Ball2015}%
  \BibitemOpen
  \bibfield  {author} {\bibinfo {author} {\bibfnamefont {H.}~\bibnamefont
  {Ball}}\ and\ \bibinfo {author} {\bibfnamefont {M.~J.}\ \bibnamefont
  {Biercuk}},\ }\bibfield  {title} {\bibinfo {title} {Walsh-synthesized noise
  filters for quantum logic},\ }\href
  {https://doi.org/10.1140/epjqt/s40507-015-0022-4} {\bibfield  {journal}
  {\bibinfo  {journal} {EPJ Quantum Technology}\ }\textbf {\bibinfo {volume}
  {2}},\ \bibinfo {pages} {11} (\bibinfo {year} {2015})}\BibitemShut {NoStop}%
\bibitem [{\citenamefont {Weidt}\ \emph {et~al.}(2016)\citenamefont {Weidt},
  \citenamefont {Randall}, \citenamefont {Webster}, \citenamefont {Lake},
  \citenamefont {Webb}, \citenamefont {Cohen}, \citenamefont {Navickas},
  \citenamefont {Lekitsch}, \citenamefont {Retzker},\ and\ \citenamefont
  {Hensinger}}]{Weidt2016}%
  \BibitemOpen
  \bibfield  {author} {\bibinfo {author} {\bibfnamefont {S.}~\bibnamefont
  {Weidt}}, \bibinfo {author} {\bibfnamefont {J.}~\bibnamefont {Randall}},
  \bibinfo {author} {\bibfnamefont {S.~C.}\ \bibnamefont {Webster}}, \bibinfo
  {author} {\bibfnamefont {K.}~\bibnamefont {Lake}}, \bibinfo {author}
  {\bibfnamefont {A.~E.}\ \bibnamefont {Webb}}, \bibinfo {author}
  {\bibfnamefont {I.}~\bibnamefont {Cohen}}, \bibinfo {author} {\bibfnamefont
  {T.}~\bibnamefont {Navickas}}, \bibinfo {author} {\bibfnamefont
  {B.}~\bibnamefont {Lekitsch}}, \bibinfo {author} {\bibfnamefont
  {A.}~\bibnamefont {Retzker}},\ and\ \bibinfo {author} {\bibfnamefont {W.~K.}\
  \bibnamefont {Hensinger}},\ }\bibfield  {title} {\bibinfo {title}
  {Trapped-ion quantum logic with global radiation fields},\ }\href
  {https://doi.org/10.1103/PhysRevLett.117.220501} {\bibfield  {journal}
  {\bibinfo  {journal} {Phys. Rev. Lett.}\ }\textbf {\bibinfo {volume} {117}},\
  \bibinfo {pages} {220501} (\bibinfo {year} {2016})}\BibitemShut {NoStop}%
\bibitem [{\citenamefont {Hahn}\ \emph {et~al.}(2019)\citenamefont {Hahn},
  \citenamefont {Zarantonello}, \citenamefont {Schulte}, \citenamefont
  {Bautista-Salvador}, \citenamefont {Hammerer},\ and\ \citenamefont
  {Ospelkaus}}]{Hahn2019a}%
  \BibitemOpen
  \bibfield  {author} {\bibinfo {author} {\bibfnamefont {H.}~\bibnamefont
  {Hahn}}, \bibinfo {author} {\bibfnamefont {G.}~\bibnamefont {Zarantonello}},
  \bibinfo {author} {\bibfnamefont {M.}~\bibnamefont {Schulte}}, \bibinfo
  {author} {\bibfnamefont {A.}~\bibnamefont {Bautista-Salvador}}, \bibinfo
  {author} {\bibfnamefont {K.}~\bibnamefont {Hammerer}},\ and\ \bibinfo
  {author} {\bibfnamefont {C.}~\bibnamefont {Ospelkaus}},\ }\bibfield  {title}
  {\bibinfo {title} {{Integrated $^9$Be$^+$ multi-qubit gate device for the
  ion-trap quantum computer}},\ }\href
  {https://doi.org/10.1038/s41534-019-0184-5} {\bibfield  {journal} {\bibinfo
  {journal} {Npj Quantum Inf.}\ }\textbf {\bibinfo {volume} {5}},\ \bibinfo
  {pages} {2} (\bibinfo {year} {2019})}\BibitemShut {NoStop}%
\bibitem [{\citenamefont {Zarantonello}\ \emph {et~al.}(2019)\citenamefont
  {Zarantonello}, \citenamefont {Hahn}, \citenamefont {Morgner}, \citenamefont
  {Schulte}, \citenamefont {Bautista-Salvador}, \citenamefont {Werner},
  \citenamefont {Hammerer},\ and\ \citenamefont
  {Ospelkaus}}]{Zarantonello2019}%
  \BibitemOpen
  \bibfield  {author} {\bibinfo {author} {\bibfnamefont {G.}~\bibnamefont
  {Zarantonello}}, \bibinfo {author} {\bibfnamefont {H.}~\bibnamefont {Hahn}},
  \bibinfo {author} {\bibfnamefont {J.}~\bibnamefont {Morgner}}, \bibinfo
  {author} {\bibfnamefont {M.}~\bibnamefont {Schulte}}, \bibinfo {author}
  {\bibfnamefont {A.}~\bibnamefont {Bautista-Salvador}}, \bibinfo {author}
  {\bibfnamefont {R.~F.}\ \bibnamefont {Werner}}, \bibinfo {author}
  {\bibfnamefont {K.}~\bibnamefont {Hammerer}},\ and\ \bibinfo {author}
  {\bibfnamefont {C.}~\bibnamefont {Ospelkaus}},\ }\bibfield  {title} {\bibinfo
  {title} {{Robust and Resource-Efficient Microwave Near-Field Entangling Be+ 9
  Gate}},\ }\href {https://doi.org/10.1103/PhysRevLett.123.260503} {\bibfield
  {journal} {\bibinfo  {journal} {Phys. Rev. Lett.}\ }\textbf {\bibinfo
  {volume} {123}},\ \bibinfo {pages} {260503} (\bibinfo {year}
  {2019})}\BibitemShut {NoStop}%
\bibitem [{\citenamefont {Duwe}\ \emph {et~al.}(2022)\citenamefont {Duwe},
  \citenamefont {Zarantonello}, \citenamefont {Pulido-Mateo}, \citenamefont
  {Mendpara}, \citenamefont {Krinner}, \citenamefont {Bautista-Salvador},
  \citenamefont {Vitanov}, \citenamefont {Hammerer}, \citenamefont {Werner},\
  and\ \citenamefont {Ospelkaus}}]{Duwe2022}%
  \BibitemOpen
  \bibfield  {author} {\bibinfo {author} {\bibfnamefont {M.}~\bibnamefont
  {Duwe}}, \bibinfo {author} {\bibfnamefont {G.}~\bibnamefont {Zarantonello}},
  \bibinfo {author} {\bibfnamefont {N.}~\bibnamefont {Pulido-Mateo}}, \bibinfo
  {author} {\bibfnamefont {H.}~\bibnamefont {Mendpara}}, \bibinfo {author}
  {\bibfnamefont {L.}~\bibnamefont {Krinner}}, \bibinfo {author} {\bibfnamefont
  {A.}~\bibnamefont {Bautista-Salvador}}, \bibinfo {author} {\bibfnamefont
  {N.~V.}\ \bibnamefont {Vitanov}}, \bibinfo {author} {\bibfnamefont
  {K.}~\bibnamefont {Hammerer}}, \bibinfo {author} {\bibfnamefont {R.~F.}\
  \bibnamefont {Werner}},\ and\ \bibinfo {author} {\bibfnamefont
  {C.}~\bibnamefont {Ospelkaus}},\ }\bibfield  {title} {\bibinfo {title}
  {Numerical optimization of amplitude-modulated pulses in microwave-driven
  entanglement generation},\ }\href {https://doi.org/10.1088/2058-9565/ac7b41}
  {\bibfield  {journal} {\bibinfo  {journal} {Quantum Science and Technology}\
  }\textbf {\bibinfo {volume} {7}},\ \bibinfo {pages} {045005} (\bibinfo {year}
  {2022})}\BibitemShut {NoStop}%
\bibitem [{\citenamefont {Weber}\ \emph {et~al.}(2023)\citenamefont {Weber},
  \citenamefont {Löschnauer}, \citenamefont {Wolf}, \citenamefont {Gely},
  \citenamefont {Hanley}, \citenamefont {Goodwin}, \citenamefont {Ballance},
  \citenamefont {Harty},\ and\ \citenamefont {Lucas}}]{Weber2022}%
  \BibitemOpen
  \bibfield  {author} {\bibinfo {author} {\bibfnamefont {M.~A.}\ \bibnamefont
  {Weber}}, \bibinfo {author} {\bibfnamefont {C.}~\bibnamefont {Löschnauer}},
  \bibinfo {author} {\bibfnamefont {J.}~\bibnamefont {Wolf}}, \bibinfo {author}
  {\bibfnamefont {M.~F.}\ \bibnamefont {Gely}}, \bibinfo {author}
  {\bibfnamefont {R.~K.}\ \bibnamefont {Hanley}}, \bibinfo {author}
  {\bibfnamefont {J.~F.}\ \bibnamefont {Goodwin}}, \bibinfo {author}
  {\bibfnamefont {C.~J.}\ \bibnamefont {Ballance}}, \bibinfo {author}
  {\bibfnamefont {T.~P.}\ \bibnamefont {Harty}},\ and\ \bibinfo {author}
  {\bibfnamefont {D.~M.}\ \bibnamefont {Lucas}},\ }\bibfield  {title} {\bibinfo
  {title} {Cryogenic ion trap system for high-fidelity near-field
  microwave-driven quantum logic},\ }\href
  {https://doi.org/10.1088/2058-9565/acfba8} {\bibfield  {journal} {\bibinfo
  {journal} {Quantum Science and Technology}\ }\textbf {\bibinfo {volume}
  {9}},\ \bibinfo {pages} {015007} (\bibinfo {year} {2023})}\BibitemShut
  {NoStop}%
\bibitem [{\citenamefont {Weber}(2022)}]{Weber2023}%
  \BibitemOpen
  \bibfield  {author} {\bibinfo {author} {\bibfnamefont {M.~A.}\ \bibnamefont
  {Weber}},\ }\href {https://doi.org/10.5287/ora-zrqaje77n} {\bibinfo {title}
  {High-fidelity, near-field microwave gates in a cryogenic surface trap}},\
  \bibinfo {howpublished} {D.Phil. thesis, University of Oxford} (\bibinfo
  {year} {2022})\BibitemShut {NoStop}%
\bibitem [{\citenamefont {Sutherland}\ \emph {et~al.}(2022)\citenamefont
  {Sutherland}, \citenamefont {Yu}, \citenamefont {Beck},\ and\ \citenamefont
  {H\"affner}}]{Sutherland2022}%
  \BibitemOpen
  \bibfield  {author} {\bibinfo {author} {\bibfnamefont {R.~T.}\ \bibnamefont
  {Sutherland}}, \bibinfo {author} {\bibfnamefont {Q.}~\bibnamefont {Yu}},
  \bibinfo {author} {\bibfnamefont {K.~M.}\ \bibnamefont {Beck}},\ and\
  \bibinfo {author} {\bibfnamefont {H.}~\bibnamefont {H\"affner}},\ }\bibfield
  {title} {\bibinfo {title} {One- and two-qubit gate infidelities due to
  motional errors in trapped ions and electrons},\ }\href
  {https://doi.org/10.1103/PhysRevA.105.022437} {\bibfield  {journal} {\bibinfo
   {journal} {Phys. Rev. A}\ }\textbf {\bibinfo {volume} {105}},\ \bibinfo
  {pages} {022437} (\bibinfo {year} {2022})}\BibitemShut {NoStop}%
\bibitem [{\citenamefont {M{\o}lmer}\ and\ \citenamefont
  {S{\o}rensen}(1999)}]{Molmer1999}%
  \BibitemOpen
  \bibfield  {author} {\bibinfo {author} {\bibfnamefont {K.}~\bibnamefont
  {M{\o}lmer}}\ and\ \bibinfo {author} {\bibfnamefont {A.}~\bibnamefont
  {S{\o}rensen}},\ }\bibfield  {title} {\bibinfo {title} {{Multiparticle
  entanglement of hot trapped ions}},\ }\href
  {https://doi.org/10.1103/PhysRevLett.82.1835} {\bibfield  {journal} {\bibinfo
   {journal} {Phys. Rev. Lett.}\ }\textbf {\bibinfo {volume} {82}},\ \bibinfo
  {pages} {1835} (\bibinfo {year} {1999})}\BibitemShut {NoStop}%
\bibitem [{\citenamefont {James}(1998)}]{James1998}%
  \BibitemOpen
  \bibfield  {author} {\bibinfo {author} {\bibfnamefont {D.~F.~V.}\
  \bibnamefont {James}},\ }\bibfield  {title} {\bibinfo {title} {Quantum
  dynamics of cold trapped ions with application to quantum computation},\
  }\href {https://doi.org/10.1007/s003400050373} {\bibfield  {journal}
  {\bibinfo  {journal} {Applied Physics B}\ }\textbf {\bibinfo {volume} {66}},\
  \bibinfo {pages} {181} (\bibinfo {year} {1998})}\BibitemShut {NoStop}%
\bibitem [{Note1()}]{Note1}%
  \BibitemOpen
  \bibinfo {note} {We neglect in Eq.~(\ref {eq:drive_Hamiltonian}) the small
  $\approx 15^\circ $ angle between the field gradient and the motional mode
  axis.}\BibitemShut {Stop}%
\bibitem [{\citenamefont {Bermudez}\ \emph {et~al.}(2012)\citenamefont
  {Bermudez}, \citenamefont {Schmidt}, \citenamefont {Plenio},\ and\
  \citenamefont {Retzker}}]{Bermudez2012}%
  \BibitemOpen
  \bibfield  {author} {\bibinfo {author} {\bibfnamefont {A.}~\bibnamefont
  {Bermudez}}, \bibinfo {author} {\bibfnamefont {P.~O.}\ \bibnamefont
  {Schmidt}}, \bibinfo {author} {\bibfnamefont {M.~B.}\ \bibnamefont
  {Plenio}},\ and\ \bibinfo {author} {\bibfnamefont {A.}~\bibnamefont
  {Retzker}},\ }\bibfield  {title} {\bibinfo {title} {Robust trapped-ion
  quantum logic gates by continuous dynamical decoupling},\ }\href
  {https://doi.org/10.1103/PhysRevA.85.040302} {\bibfield  {journal} {\bibinfo
  {journal} {Phys. Rev. A}\ }\textbf {\bibinfo {volume} {85}},\ \bibinfo
  {pages} {040302} (\bibinfo {year} {2012})}\BibitemShut {NoStop}%
\bibitem [{\citenamefont {Leibfried}\ \emph {et~al.}(2003)\citenamefont
  {Leibfried}, \citenamefont {DeMarco}, \citenamefont {Meyer}, \citenamefont
  {Lucas}, \citenamefont {Barrett}, \citenamefont {Britton}, \citenamefont
  {Itano}, \citenamefont {Jelenkovi{\'{c}}}, \citenamefont {Langer},
  \citenamefont {Rosenband},\ and\ \citenamefont {Wineland}}]{Leibfried2003}%
  \BibitemOpen
  \bibfield  {author} {\bibinfo {author} {\bibfnamefont {D.}~\bibnamefont
  {Leibfried}}, \bibinfo {author} {\bibfnamefont {B.}~\bibnamefont {DeMarco}},
  \bibinfo {author} {\bibfnamefont {V.}~\bibnamefont {Meyer}}, \bibinfo
  {author} {\bibfnamefont {D.}~\bibnamefont {Lucas}}, \bibinfo {author}
  {\bibfnamefont {M.}~\bibnamefont {Barrett}}, \bibinfo {author} {\bibfnamefont
  {J.}~\bibnamefont {Britton}}, \bibinfo {author} {\bibfnamefont {W.~M.}\
  \bibnamefont {Itano}}, \bibinfo {author} {\bibfnamefont {B.}~\bibnamefont
  {Jelenkovi{\'{c}}}}, \bibinfo {author} {\bibfnamefont {C.}~\bibnamefont
  {Langer}}, \bibinfo {author} {\bibfnamefont {T.}~\bibnamefont {Rosenband}},\
  and\ \bibinfo {author} {\bibfnamefont {D.~J.}\ \bibnamefont {Wineland}},\
  }\bibfield  {title} {\bibinfo {title} {{Experimental demonstration of a
  robust, high-fidelity geometric two ion-qubit phase gate}},\ }\href
  {https://doi.org/10.1038/nature01492} {\bibfield  {journal} {\bibinfo
  {journal} {Nature}\ }\textbf {\bibinfo {volume} {422}},\ \bibinfo {pages}
  {412} (\bibinfo {year} {2003})}\BibitemShut {NoStop}%
\bibitem [{\citenamefont {Nie}\ \emph {et~al.}(2009)\citenamefont {Nie},
  \citenamefont {Roos},\ and\ \citenamefont {James}}]{Nie2009}%
  \BibitemOpen
  \bibfield  {author} {\bibinfo {author} {\bibfnamefont {X.~R.}\ \bibnamefont
  {Nie}}, \bibinfo {author} {\bibfnamefont {C.~F.}\ \bibnamefont {Roos}},\ and\
  \bibinfo {author} {\bibfnamefont {D.~F.}\ \bibnamefont {James}},\ }\bibfield
  {title} {\bibinfo {title} {Theory of cross phase modulation for the
  vibrational modes of trapped ions},\ }\href
  {https://doi.org/https://doi.org/10.1016/j.physleta.2008.11.045} {\bibfield
  {journal} {\bibinfo  {journal} {Phys. Lett. A}\ }\textbf {\bibinfo {volume}
  {373}},\ \bibinfo {pages} {422} (\bibinfo {year} {2009})}\BibitemShut
  {NoStop}%
\bibitem [{\citenamefont {Gaebler}\ \emph {et~al.}(2012)\citenamefont
  {Gaebler}, \citenamefont {Meier}, \citenamefont {Tan}, \citenamefont
  {Bowler}, \citenamefont {Lin}, \citenamefont {Hanneke}, \citenamefont {Jost},
  \citenamefont {Home}, \citenamefont {Knill}, \citenamefont {Leibfried},\ and\
  \citenamefont {Wineland}}]{Gaebler2012}%
  \BibitemOpen
  \bibfield  {author} {\bibinfo {author} {\bibfnamefont {J.~P.}\ \bibnamefont
  {Gaebler}}, \bibinfo {author} {\bibfnamefont {A.~M.}\ \bibnamefont {Meier}},
  \bibinfo {author} {\bibfnamefont {T.~R.}\ \bibnamefont {Tan}}, \bibinfo
  {author} {\bibfnamefont {R.}~\bibnamefont {Bowler}}, \bibinfo {author}
  {\bibfnamefont {Y.}~\bibnamefont {Lin}}, \bibinfo {author} {\bibfnamefont
  {D.}~\bibnamefont {Hanneke}}, \bibinfo {author} {\bibfnamefont {J.~D.}\
  \bibnamefont {Jost}}, \bibinfo {author} {\bibfnamefont {J.~P.}\ \bibnamefont
  {Home}}, \bibinfo {author} {\bibfnamefont {E.}~\bibnamefont {Knill}},
  \bibinfo {author} {\bibfnamefont {D.}~\bibnamefont {Leibfried}},\ and\
  \bibinfo {author} {\bibfnamefont {D.~J.}\ \bibnamefont {Wineland}},\
  }\bibfield  {title} {\bibinfo {title} {Randomized benchmarking of multiqubit
  gates},\ }\href {https://doi.org/10.1103/PhysRevLett.108.260503} {\bibfield
  {journal} {\bibinfo  {journal} {Phys. Rev. Lett.}\ }\textbf {\bibinfo
  {volume} {108}},\ \bibinfo {pages} {260503} (\bibinfo {year}
  {2012})}\BibitemShut {NoStop}%
\bibitem [{\citenamefont {Gerster}\ \emph {et~al.}(2022)\citenamefont
  {Gerster}, \citenamefont {Mart\'{\i}nez-Garc\'{\i}a}, \citenamefont {Hrmo},
  \citenamefont {van Mourik}, \citenamefont {Wilhelm}, \citenamefont {Vodola},
  \citenamefont {M\"uller}, \citenamefont {Blatt}, \citenamefont {Schindler},\
  and\ \citenamefont {Monz}}]{Gerster2022}%
  \BibitemOpen
  \bibfield  {author} {\bibinfo {author} {\bibfnamefont {L.}~\bibnamefont
  {Gerster}}, \bibinfo {author} {\bibfnamefont {F.}~\bibnamefont
  {Mart\'{\i}nez-Garc\'{\i}a}}, \bibinfo {author} {\bibfnamefont
  {P.}~\bibnamefont {Hrmo}}, \bibinfo {author} {\bibfnamefont {M.~W.}\
  \bibnamefont {van Mourik}}, \bibinfo {author} {\bibfnamefont
  {B.}~\bibnamefont {Wilhelm}}, \bibinfo {author} {\bibfnamefont
  {D.}~\bibnamefont {Vodola}}, \bibinfo {author} {\bibfnamefont
  {M.}~\bibnamefont {M\"uller}}, \bibinfo {author} {\bibfnamefont
  {R.}~\bibnamefont {Blatt}}, \bibinfo {author} {\bibfnamefont
  {P.}~\bibnamefont {Schindler}},\ and\ \bibinfo {author} {\bibfnamefont
  {T.}~\bibnamefont {Monz}},\ }\bibfield  {title} {\bibinfo {title}
  {Experimental bayesian calibration of trapped-ion entangling operations},\
  }\href {https://doi.org/10.1103/PRXQuantum.3.020350} {\bibfield  {journal}
  {\bibinfo  {journal} {PRX Quantum}\ }\textbf {\bibinfo {volume} {3}},\
  \bibinfo {pages} {020350} (\bibinfo {year} {2022})}\BibitemShut {NoStop}%
\bibitem [{\citenamefont {Hayes}\ \emph {et~al.}(2012)\citenamefont {Hayes},
  \citenamefont {Clark}, \citenamefont {Debnath}, \citenamefont {Hucul},
  \citenamefont {Inlek}, \citenamefont {Lee}, \citenamefont {Quraishi},\ and\
  \citenamefont {Monroe}}]{Hayes2012}%
  \BibitemOpen
  \bibfield  {author} {\bibinfo {author} {\bibfnamefont {D.}~\bibnamefont
  {Hayes}}, \bibinfo {author} {\bibfnamefont {S.~M.}\ \bibnamefont {Clark}},
  \bibinfo {author} {\bibfnamefont {S.}~\bibnamefont {Debnath}}, \bibinfo
  {author} {\bibfnamefont {D.}~\bibnamefont {Hucul}}, \bibinfo {author}
  {\bibfnamefont {I.~V.}\ \bibnamefont {Inlek}}, \bibinfo {author}
  {\bibfnamefont {K.~W.}\ \bibnamefont {Lee}}, \bibinfo {author} {\bibfnamefont
  {Q.}~\bibnamefont {Quraishi}},\ and\ \bibinfo {author} {\bibfnamefont
  {C.}~\bibnamefont {Monroe}},\ }\bibfield  {title} {\bibinfo {title} {Coherent
  error suppression in multiqubit entangling gates},\ }\href
  {https://doi.org/10.1103/PhysRevLett.109.020503} {\bibfield  {journal}
  {\bibinfo  {journal} {Phys. Rev. Lett.}\ }\textbf {\bibinfo {volume} {109}},\
  \bibinfo {pages} {020503} (\bibinfo {year} {2012})}\BibitemShut {NoStop}%
\bibitem [{\citenamefont {H\"affner}\ \emph {et~al.}(2003)\citenamefont
  {H\"affner}, \citenamefont {Gulde}, \citenamefont {Riebe}, \citenamefont
  {Lancaster}, \citenamefont {Becher}, \citenamefont {Eschner}, \citenamefont
  {Schmidt-Kaler},\ and\ \citenamefont {Blatt}}]{Haffner2003}%
  \BibitemOpen
  \bibfield  {author} {\bibinfo {author} {\bibfnamefont {H.}~\bibnamefont
  {H\"affner}}, \bibinfo {author} {\bibfnamefont {S.}~\bibnamefont {Gulde}},
  \bibinfo {author} {\bibfnamefont {M.}~\bibnamefont {Riebe}}, \bibinfo
  {author} {\bibfnamefont {G.}~\bibnamefont {Lancaster}}, \bibinfo {author}
  {\bibfnamefont {C.}~\bibnamefont {Becher}}, \bibinfo {author} {\bibfnamefont
  {J.}~\bibnamefont {Eschner}}, \bibinfo {author} {\bibfnamefont
  {F.}~\bibnamefont {Schmidt-Kaler}},\ and\ \bibinfo {author} {\bibfnamefont
  {R.}~\bibnamefont {Blatt}},\ }\bibfield  {title} {\bibinfo {title} {Precision
  measurement and compensation of optical stark shifts for an ion-trap quantum
  processor},\ }\href {https://doi.org/10.1103/PhysRevLett.90.143602}
  {\bibfield  {journal} {\bibinfo  {journal} {Phys. Rev. Lett.}\ }\textbf
  {\bibinfo {volume} {90}},\ \bibinfo {pages} {143602} (\bibinfo {year}
  {2003})}\BibitemShut {NoStop}%
\bibitem [{\citenamefont {Smith}\ \emph {et~al.}(2023)\citenamefont {Smith},
  \citenamefont {Leu}, \citenamefont {Gely},\ and\ \citenamefont
  {Lucas}}]{Smith2023}%
  \BibitemOpen
  \bibfield  {author} {\bibinfo {author} {\bibfnamefont {M.~C.}\ \bibnamefont
  {Smith}}, \bibinfo {author} {\bibfnamefont {A.~D.}\ \bibnamefont {Leu}},
  \bibinfo {author} {\bibfnamefont {M.~F.}\ \bibnamefont {Gely}},\ and\
  \bibinfo {author} {\bibfnamefont {D.~M.}\ \bibnamefont {Lucas}},\ }\bibfield
  {title} {\bibinfo {title} {Focusing of quantum gate interactions using
  dynamical decoupling},\ }\href {https://arxiv.org/abs/2309.02125} {\bibfield
  {journal} {\bibinfo  {journal} {arXiv preprint arXiv:2309.02125}\ } (\bibinfo
  {year} {2023})}\BibitemShut {NoStop}%
\bibitem [{\citenamefont {Nünnerich}\ \emph {et~al.}(2024)\citenamefont
  {Nünnerich}, \citenamefont {Cohen}, \citenamefont {Barthel}, \citenamefont
  {Huber}, \citenamefont {Niroomand}, \citenamefont {Retzker},\ and\
  \citenamefont {Wunderlich}}]{nunnerich2024fast}%
  \BibitemOpen
  \bibfield  {author} {\bibinfo {author} {\bibfnamefont {M.}~\bibnamefont
  {Nünnerich}}, \bibinfo {author} {\bibfnamefont {D.}~\bibnamefont {Cohen}},
  \bibinfo {author} {\bibfnamefont {P.}~\bibnamefont {Barthel}}, \bibinfo
  {author} {\bibfnamefont {P.~H.}\ \bibnamefont {Huber}}, \bibinfo {author}
  {\bibfnamefont {D.}~\bibnamefont {Niroomand}}, \bibinfo {author}
  {\bibfnamefont {A.}~\bibnamefont {Retzker}},\ and\ \bibinfo {author}
  {\bibfnamefont {C.}~\bibnamefont {Wunderlich}},\ }\href@noop {} {\bibinfo
  {title} {Fast, robust and laser-free universal entangling gates for
  trapped-ion quantum computing}} (\bibinfo {year} {2024}),\ \Eprint
  {https://arxiv.org/abs/2403.04730} {arXiv:2403.04730} \BibitemShut {NoStop}%
\bibitem [{\citenamefont {Löschnauer}\ \emph {et~al.}(2024)\citenamefont
  {Löschnauer}, \citenamefont {Toba}, \citenamefont {Hughes}, \citenamefont
  {King}, \citenamefont {Weber}, \citenamefont {Srinivas}, \citenamefont
  {Matt}, \citenamefont {Nourshargh}, \citenamefont {Allcock}, \citenamefont
  {Ballance}, \citenamefont {Matthiesen}, \citenamefont {Malinowski},\ and\
  \citenamefont {Harty}}]{loschnauer2024}%
  \BibitemOpen
  \bibfield  {author} {\bibinfo {author} {\bibfnamefont {C.~M.}\ \bibnamefont
  {Löschnauer}}, \bibinfo {author} {\bibfnamefont {J.~M.}\ \bibnamefont
  {Toba}}, \bibinfo {author} {\bibfnamefont {A.~C.}\ \bibnamefont {Hughes}},
  \bibinfo {author} {\bibfnamefont {S.~A.}\ \bibnamefont {King}}, \bibinfo
  {author} {\bibfnamefont {M.~A.}\ \bibnamefont {Weber}}, \bibinfo {author}
  {\bibfnamefont {R.}~\bibnamefont {Srinivas}}, \bibinfo {author}
  {\bibfnamefont {R.}~\bibnamefont {Matt}}, \bibinfo {author} {\bibfnamefont
  {R.}~\bibnamefont {Nourshargh}}, \bibinfo {author} {\bibfnamefont {D.~T.~C.}\
  \bibnamefont {Allcock}}, \bibinfo {author} {\bibfnamefont {C.~J.}\
  \bibnamefont {Ballance}}, \bibinfo {author} {\bibfnamefont {C.}~\bibnamefont
  {Matthiesen}}, \bibinfo {author} {\bibfnamefont {M.}~\bibnamefont
  {Malinowski}},\ and\ \bibinfo {author} {\bibfnamefont {T.~P.}\ \bibnamefont
  {Harty}},\ }\bibfield  {title} {\bibinfo {title} {Scalable, high-fidelity
  all-electronic control of trapped-ion qubits},\ }\href
  {https://arxiv.org/abs/2407.07694} {\bibfield  {journal} {\bibinfo  {journal}
  {arXiv preprint arXiv:2407.07694}\ } (\bibinfo {year} {2024})}\BibitemShut
  {NoStop}%
\end{thebibliography}%

\end{document}